\documentclass[symmetry,article,submit,moreauthors,pdftex]{mdpi}

\firstpage{1}
\pubvolume{xx}
\issuenum{1}
\articlenumber{5}
\pubyear{2019}
\copyrightyear{2019}
\history{Received: date; Accepted: date; Published: date}

\usepackage{epsfig,amsfonts,mathrsfs}
\usepackage{array}
\usepackage{diagbox}
\usepackage{enumitem}
\usepackage{amsmath,amssymb}
\usepackage{bm}
\usepackage{longtable}
\usepackage[T1]{fontenc}
\usepackage[utf8]{inputenc}
\usepackage{verbatim}
\usepackage{pifont}
\usepackage{yfonts}
\usepackage{url}
\usepackage{multirow}
\usepackage{tensor}
\usepackage{mathtools}
\usepackage[normalem]{ulem}
\usepackage{fancyhdr}
\usepackage{xcolor}
\usepackage{soul}
\usepackage{amssymb,amsmath,verbatim,mathtools,needspace,enumitem,etoolbox,graphicx,physics,microtype,afterpage,dcolumn,diagbox}

\definecolor{linkcolor}{rgb}{0.0,0.3,0.5}
\def\dbar{{~~\mathchar'26\mkern-11mu d}}

\usepackage[T1]{fontenc}
\usepackage[utf8]{inputenc}
\usepackage{tabularx}
\allowdisplaybreaks

\usepackage{lipsum}

\usepackage{diagbox}

\usepackage{marvosym}
\usepackage{yfonts}
\usepackage{upgreek}
\usepackage{pdflscape}

\Title{Structure of neutron stars in massive scalar-tensor gravity}

\newcommand{\myorcid}[1]{{\href{https://orcid.org/#1}{\includegraphics[width=0.32cm]{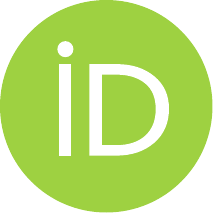}}}}

\Author{
Roxana Rosca-Mead$^{1}$ \myorcid{0000-0001-5666-1033},
Christopher J. Moore$^{2}$ \myorcid{0000-0002-2527-0213},
Ulrich Sperhake$^{1,3}$ \myorcid{0000-0002-3134-7088},
Michalis Agathos$^{1,4}$ \myorcid{0000-0002-9072-1121}
and
Davide Gerosa$^{2}$ \myorcid{0000-0002-0933-3579}
}

\AuthorNames{Roxana Rosca-Mead, Christopher J. Moore, Ulrich Sperhake, Michalis Agathos and Davide Gerosa}

\address{%
$^{1}$ \quad DAMTP, Centre for Mathematical Sciences, University of Cambridge, Wilberforce Road, Cambridge CB3 0WA, UK\\
$^{2}$ \quad School of Physics and Astronomy \& Institute for Gravitational Wave Astronomy, University of Birmingham, Birmingham, B15 2TT, UK\\
$^{3}$ \quad TAPIR 350-17, California Institute of Technology,
1200 E. California Blvd., Pasadena, California 91125, USA\\
$^{4}$ \quad Kavli Institute for Cosmology Cambridge, Madingley Road Cambridge CB3 0HA, UK}

\corres{Correspondence: rr417@cam.ac.uk (R.R)}

\abstract{
We compute families of spherically symmetric neutron-star
models in two-derivative scalar-tensor theories of gravity
with a massive scalar field. The numerical approach
we present allows us to compute the
resulting spacetimes out to infinite radius using a
relaxation algorithm on a compactified grid. We discuss
the structure of the weakly and strongly scalarized
branches of neutron-star models thus obtained and their
dependence on the linear and quadratic coupling parameters
$\alpha_0$, $\beta_0$
between the scalar and tensor sectors of the theory, as well
as the scalar mass $\mu$. For highly negative values of
$\beta_0$, we encounter configurations resembling a ``gravitational atom'', 
consisting of a highly compact baryon star surrounded by
a scalar cloud. A stability analysis based on binding-energy
calculations suggests that these configurations are unstable
and we expect them to migrate to models with radially decreasing
baryon density {\it and} scalar field strength.
}

\keyword{Modified gravity, Scalar-tensor theory, Compact objects, Relativistic astrophysics}

\begin{document}

\section{Introduction}
\label{sec:intro}
Ever since its formulation in 1915, Einstein's general relativity (GR)
has been a tremendously successful theory of gravity, combining
mathematical elegance with enormous predictive power. Phenomena
ranging from Mercury's perihelion precession to
the formation of black holes (BHs),
the generation of gravitational waves
and the big bang, find a mathematical
description within this single theory.
A wide range of lab-based experiments, solar-system tests and observations of astronomical phenomena have systematically scrutinized the
accuracy of the theory's predictions and unanimously seen GR
passing these tests with flying colors~\cite{Will:2014kxa}.
With the advent of gravitational-wave (GW) astronomy, marked by the detection of GW150914 by LIGO~\cite{Abbott:2016blz},
new tests of GR have become possible in spacetime regions with strong and dynamical gravitational fields and sources moving at relativistic velocities.
Once again, all GW observations so far are
compatible with GR \cite{TheLIGOScientific:2016src,Monitor:2017mdv,Abbott:2018lct,LIGOScientific:2019fpa,Yunes:2016jcc}.

Notwithstanding GR's success,
the search for possible alternative theories of gravity has for many
decades been a highly active area of research
\cite{Clifton:2011jh,Will:2014kxa,Berti:2015itd,Barack:2018yly,Olmo:2019flu,Capozziello:2019cav}, motivated by
important theoretical considerations, such as the incompatibility of GR with quantum theory at a fundamental level,
as well as open questions in observational astronomy and cosmology.
Astronomical observations of galactic rotation curves, micro-lensing,
primordial nucleosynthesis or the accelerated expansion of the
universe cannot be explained in GR without evoking {\em dark matter}
and {\em dark energy}, enigmatic entities beyond the standard model of particles;
see e.g.~\cite{Freese:2008cz,Novosyadlyj:2015zpa}.

Alternatively to either the dark-matter or dark-energy
hypotheses, we may consider modifications in the laws
of gravity; just in the same way GR explained Mercury's
anomalous perihelion precession in terms of modifications of the
then prevailing Newtonian laws of gravity.
Modifications of GR may also overcome one of the most
important theoretical concerns about Einstein's theory, its
nonrenormalizability in quantum theory terms \cite{Hamber:2009zz}.
For a theory as well established
as GR, however, the quest for modifications
faces an obvious difficulty; the longstanding success of
the old theory suggests that modifications either be
extremely weak or become measurable only under new, in some sense
extreme, conditions. Quite remarkably, however, this conclusion
is not quite correct: nonperturbative effects of an alternative
theory of gravity may lead to order-of-unity deviations from
GR even if departures at {\em linearized} level
are small. The prototypical example of this phenomenon is
the {\em spontaneous scalarization} of neutron stars (NSs)
\cite{Lasky:2015uia,Ozel:2016oaf}
in scalar-tensor (ST) theory of gravity discovered by
Damour and Esposito-Far{\`e}se in 1993 \cite{Damour:1993hw}.
Here, the additional degree of freedom -- in the form of
the scalar field -- allows for additional families of
solutions describing stars in equilibrium. Moreover, these
new families of solutions may appear ``abruptly'', in a manner
akin to phase-transitions, as one varies certain parameters of the
theory or the star's density profile.
In the case of compact stars in ST gravity, the new solutions
consist of stars with strong
scalar-field profiles, as opposed to the GR-like models with
negligible or zero scalar field. Often, the new scalarized
configurations are energetically favored over their GR-like counterparts
(assuming equal baryon mass or number), so that they
represent the
expected endpoints of dynamical scenarios.

Spontaenous scalarization bears a qualitative resemblence to
other effects known in physics; Damour and Esposito-Far{\`e}se
have highlighted its analogy to the spontaneous magnetization
of ferromagnets \cite{Damour:1996ke} and later studies have
interpreted its onset in terms of catastrophe theory
\cite{Harada:1998ge} or a tachyonic instability
\cite{Ramazanoglu:2016kul}.
Originally, spontaneous scalarization has been identified for
spherically symmetric NS models in a class of massless
ST theories sometimes refered to as Bergmann-Wagoner
\cite{Bergmann:1968ve,Wagoner:1970vr} theories; these complement the
metric sector of GR with a single scalar field, are governed by second-order
covariant field equations at most linear in second and quadratic
in first derivatives, and obey the weak equivalence principle.
The phenomenon has by now been demonstrated to occur over a wide
range of configurations and also in other theoretical frameworks~\cite{Andreou:2019ikc,Ventagli:2020rnx}.
Analogous phenomena occur in many scenarios involving fields
non-minimally coupled to a spacetime metric.  Examples include
scalarized BHs in theories with Gauss-Bonnet coupling~\cite{Silva:2017uqg,Doneva:2017bvd},
universal horizons in Lorentz violating gravity theories~\cite{Barausse:2011pu}
and the spotaneous {\em vectorization} or {\em tensorization}
of compact stars in modified gravity~\cite{Ramazanoglu:2017xbl,Annulli:2019fzq,Ramazanoglu:2019gbz}.

Numerous studies have demonstrated that spontaneous scalarization
features as prominently in rotating NS models,
either in the slow-rotation limit
\cite{Sotani:2012eb,Pani:2014jra,Silva:2014fca,Yazadjiev:2016pcb,Motahar:2017blm,Staykov:2018hhc},
for fast rotation
\cite{Doneva:2013qva,Doneva:2014faa},
or with differential rotation
\cite{Doneva:2018ouu}.
Spontaneous scalarization has also been found a robust phenomenon
under variations of the equation of state (EOS)
\cite{Ramazanoglu:2016kul,Motahar:2017blm,Sotani:2017pfj,Rezaei:2018jur,Anderson:2019hio}.
While quantitative differences occur, the phenomenon as such appears
to be ubiquitous and also preserve the approximate
EOS universality of the relation between the moment of inertia $I$
and the quadrupole moment $Q$ known from GR
\cite{Doneva:2014faa,Pani:2014jra} and the inertia vs.~compactness
universality \cite{Motahar:2017blm}.

Numerical calculations find that the families of scalarized NSs can have larger maximal masses than
the corresponding GR solutions~\cite{Damour:1993hw,Ramazanoglu:2016kul,Morisaki:2017nit,Doneva:2018ouu}.
Often this is accompanied with an even stronger increase in the NS radius,
so that the maximum compactness of NSs in ST gravity is smaller than in GR~\cite{Sotani:2018aiz}.
These findings suggest that the scalar field may
effectively stiffen the equation of state and thus
counteract the normal gravitational pull. This effect, does
not appear to be generic, however, but rather depends on details of the
matter sources. A generalization of the Buchdahl limit
\cite{Tsuchida:1998jw} has found that compactness above the Buchdahl limit
is possible in ST theory, but only if the energy density
$\rho$ and pressure $p$ satisfy $\rho < 3p$.
Sotani and Kokkotas \cite{Sotani:2017pfj} find that the maximum NS mass
is larger in ST theory than in GR for sufficiently small sound speeds in
the core, but that the reverse holds if this velocity exceeds 0.79 of the
speed of light.
In light of the recent discovery by LIGO and Virgo of the compact binary GW190814,
whose light component's mass likely falls in the so-called mass gap between NS and BHs~\cite{Abbott:2020khf},
it is worth noting that massive ST gravity allows for the possibility of such objects
being strongly scalarized NSs.

The presence or absence of the spontaneous scalarization phenomenon
is largely determined by the quadratic coupling parameter
$\beta_0$ between the
scalar and tensor sectors of the theory; cf.~Eq.(\ref{eq:conformalfactor}) below. Strongly
scalarized NS models are found for $\beta_0 \lesssim -4.35$ and this
threshold has been found remarkably robust against variations
of other parameters such as the EOS; see e.g.~\cite{Novak:1998rk,Pani:2014jra}.
In a series of studies, however, Mendes and collaborators
\cite{Mendes:2014vna,Mendes:2014ufa,Mendes:2016fby}
have demonstrated that strongly scalarized solutions can also be
obtained for positive values of $\beta_0$ {\em provided} there exist
stable equilibrium solutions for matter fields in the GR limit where the
trace $T$ of the energy-momentum tensor acquires positive values.
This can be understood, for instance, in terms of the tachyonic
instability by noticing that the scalar field is sourced by a
term $\propto \beta_0 T$; cf.~Eq.~(3) in Ref.~\cite{Ramazanoglu:2016kul}.
This $\beta_0 > 0$ scenario has been explored in time evolutions
of NSs close to the upper NS mass limit in Ref.~\cite{Palenzuela:2015ima}.
These simulations demonstrate an instability of the star to collapse
for large $\beta_0$ of $\mathcal{O}(10^2)$, suggesting an upper bound
on the parameter $\beta_0$.

Massless ST theories of gravity have by now been significantly constrained --
not least of all because of the large magnitude of the spontaneous
scalarization effect -- by the Cassini mission~\cite{Bertotti:2003rm},
Lunar Laser Ranging~\cite{Williams:2005rv},
binary pulsar observations~\cite{Wex:2014nva}
and gravitational wave (GW) observations with LIGO-Virgo~\cite{Yunes:2016jcc}.
While spontaneous scalarization has been seen to occur
in dynamical evolutions
in massless ST theory, either for the gravitational
collapse of single stars~\cite{Novak:1997hw,Novak:1998rk,Novak:1999jg,Gerosa:2016fri}
or the merger of binary NSs~\cite{Barausse:2012da,Palenzuela:2013hsa,Shibata:2013pra},
the most recent constraints on $\beta_0$ severely limit the magnitude
of the resulting GW signals and, thus, make it difficult to constrain
this theory further with GW observations.

In the context of this work, the most important extension of
the scalarization phenomenon is the inclusion of a non-zero
scalar mass.
This is because the above constraints only apply to ST
theories with a scalar mass parameter $\mu \lesssim 10^{-16}\,{\rm eV}$.
Otherwise, the Compton wavelength $\lambda_c=(2\pi \hbar)/(\mu c)$
is smaller than the distance between the objects involved in the
systems under consideration and,
hence, the scalar contribution to the objects' interaction
is suppressed. 
GW observations, on the other hand, provide exquisite constraints on dispersion which in turn can be interpreted as a constraint on the graviton mass, but this does not apply to radiation in the scalar sector.
In consequence, massive ST theory
remains compatible with present observations over much of its
parameter space.

Motivated by this realization, many recent studies have explored
spontaneous scalarization in massive ST gravity. Computations
of stationary models have confirmed that the spontaneous
scalarization phenomenon persists under the inclusion of
a mass term in the scalar potential
\cite{Chen:2015zmx,Doneva:2016xmf,Yazadjiev:2016pcb,Ramazanoglu:2017xbl,Morisaki:2017nit}.
A non-zero scalar mass $\mu > 0$ does, however, dramatically
affect the GW signals generated in stellar collapse in ST gravity
through dispersion. A Fourier mode with frequency $\omega$ propagates
at group velocity $v_g = \sqrt{1-\omega_*^2/\omega^2}$,
$\omega_*=c^2 \mu/\hbar$, so that high-frequencies reach a detector
first with lower frequencies arriving later, so the signal acquires an {\em inverse-chirp} or {\em howl}
character.
Furthermore these signals get extremely stretched out and become approximately
monochromatic (in the sense that the frequency changes by very little over one period; $\mathrm{d}f/\mathrm{dt}\ll f^2$),
can reach considerable amplitude for sufficiently negative
$\beta_0$ and may last for years or even centuries for
scalar masses $\mu \lesssim 10^{-12}~{\rm eV}$
\cite{Sperhake:2017itk,Rosca-Mead:2020ehn,Geng:2020slq}. While the inclusion
of self-interaction terms may reduce the degree of scalarization
and the amplitude of the scalar GWs
\cite{Cheong:2018gzn,Staykov:2018hhc}, this requires considerable finetuning
of the scalar potential parameters
\cite{Rosca-Mead:2019seq}.

In this work, we focus on spherically symmetric, static NS models
in the framework of Bergmann-Wagoner ST theory with massive scalar
fields. The main purpose of our study is two-fold. First, to introduce
a numerical scheme that enables us to compute these stellar models
over the entire spatial domain, all the way out to infinity, while
maintaining complete control over exponentially diverging solutions.
Second, to present an in-depth analysis of the structure of the
different solution branches and their dependence on the parameters
of the ST theory. We begin this discussion in
Sec.~\ref{sec:formalism} with a review
of the field equations governing the stars.
In Sec.~\ref{sec:numerics}, we describe the numerical
framework used for our computations. Our results on the structure
of NS solutions in massive ST gravity are presented in
Sec.~\ref{results} and we conclude in
Sec.~\ref{sec:conclusions}.

\section{Formalism}
\label{sec:formalism}
The Bergmann-Wagoner class of ST theories, i.e.~theories involving a single
scalar field that are governed by two-derivative, covariant field equations
and obey the weak equivalence principle, can be described in terms
of the action
\cite{Fujii:2003pa}
\begin{equation}
  S_J = \int dx^4 \sqrt{-g} \left[ \frac{F(\phi)}{16\pi G} R
        - \frac{1}{2}Z(\phi)
        g^{\mu \nu} (\partial_{\mu} \phi)(\partial_{\nu}\phi)
        - W(\phi) \right] + S_m(\psi_m,g_{\mu \nu})\,.
  \label{eq:Jordan_action}
\end{equation}
Here, $\psi_m$ collectively denotes the matter fields
and $S_m$ represents their coupling to the spacetime geometry
of the {\em physical} or {\em Jordan} metric $g_{\mu\nu}$
with determinant $g$ and Ricci scalar $R$.
The functions $F(\phi)$ and $Z(\phi)$ encapsulate the nonminimal coupling
of the scalar field $\phi$ to the metric sector, and $V(\phi)$
is the potential function. As we shall see shortly, the function $Z$
can be eliminated through an appropriate redefinition and is therefore
often set to unity in the literature; see e.g.~\cite{Salgado:2005hx}.

This class of theories is conveniently described in the so-called Einstein frame, obtained from the physical or Jordan frame through a conformal transformation of the metric and a redefinition of the scalar degree of freedom and its potential,
\begin{equation}
  g_{\alpha \beta}=\frac{1}{F}\bar{g}_{\alpha \beta},~~~~~~~~~~
  \frac{\partial \varphi}{\partial \phi}
        = \sqrt{\frac{3}{4} \frac{F_{,\phi}(\phi)^2}{F(\phi)^2}
        + \frac{4\pi Z(\phi)}{F(\phi)}} \,,~~~~~~~~~~
  V(\varphi) = \frac{4\pi W(\phi)}{F(\phi)^2}\,,
  \label{eq:EJtrafo}
\end{equation}
where $F_{,\phi}=dF/d\phi$.
In terms of these new functions, the action (\ref{eq:Jordan_action}) becomes
\begin{equation}
  S_E = \frac{1}{16\pi G} \int dx^4 \sqrt{-\bar{g}} \left[
        \bar{R} - 2\bar{g}^{\mu \nu} (\partial_{\mu} \varphi)
        (\partial_{\nu} \varphi) - 4V(\varphi) \right]
        +S_m\left[ \psi_m,\frac{\bar{g}_{\mu \nu}}{F(\varphi)}\right]\,,
  \label{eq: einsteinaction}
\end{equation}
where an overbar distinguishes tensors in the Einstein frame from
their Jordan counterparts.
Henceforth, we use natural units where
$c=G=1$, unless stated otherwise.

By transforming to the Einstein frame, we have eliminated the function $Z$.
The equivalence (or lack thereof) of the Einstein and Jordan frame
formulations has been the subject of a long standing debate
(see e.g.~\cite{Faraoni:1999hp,Salgado:2008xh,Geng:2020ftu} and
references therein). Without entering this debate here, we merely note
the extra freedom that the function $Z$ introduces to the transformation
(\ref{eq:EJtrafo}) between the frames and henceforth follow the recommendation
of Ref.~\cite{Faraoni:1999hp} and work in the Einstein frame.

To complete the description of the gravitational theory we must specify the remaining free functions $F$ and $V$.
Following most previous studies in the literature, we write the conformal factor
as\footnote{We note that alternative, equivalent formulations
use instead the function
$A=F^{-2}$ and/or replace $\alpha_0$ in terms of a non-zero
asymptotic value $\varphi_0$; cf.~the discussion in Sec.~3.2 of
Ref.~\cite{Gerosa:2016fri}.}
\begin{equation}
  F(\varphi) = e^{-2\alpha_0 \varphi-\beta_0 \varphi^2} \,,
  \label{eq:conformalfactor}
\end{equation}
and take as our potential function the quadratic function
\begin{equation}
  V(\varphi) = \frac{\mu^2\varphi^2}{2\hbar^2}\,,
\end{equation}
which describes a non-self-interacting scalar field of mass $\mu$.

The field equations obtained by varying the Einstein
action~(\ref{eq: einsteinaction}) with respect to
$\bar{g}_{\alpha\beta}$, $\varphi$ are given by
\begin{eqnarray}
  \bar{R}_{\alpha\beta}-\frac{1}{2}\bar{g}_{\alpha\beta} \bar{R} &=& 2\partial_{\alpha}\varphi \partial_{\beta}\varphi
        -\bar{g}_{\alpha\beta}\bar{g}^{\mu\nu}\partial_{\mu}\varphi
         \partial_{\nu}\varphi+8\pi \bar{T}_{\alpha\beta}
         -2V\bar{g}_{\alpha\beta}  \label{eq:barG}, \\
  \bar{\Box}\varphi &=& 2\pi
        \frac{F_{,\varphi}}{F} \bar{T}+V_{,\varphi}\,,
  \label{eq:boxvarphi}
\end{eqnarray}
with the energy momentum tensor
\begin{eqnarray}
  &&\bar{T}^{\alpha\beta} = \frac{2}{\sqrt{-\bar{g}}}
        \frac{\delta S_m}{\delta \bar{g}_{\alpha\beta}}
        = \frac{1}{{F(\varphi)}^3}\frac{2}{\sqrt{-g}}
        \frac{\delta S_m}{\delta g_{\alpha\beta}}
        = \frac{1}{{F(\varphi)}^3}T^{\alpha\beta}\,,
\end{eqnarray}
for which the Bianchi identity now implies the following conservation law
\begin{equation}
        \bar{\nabla}_{\mu}\bar{T}^{\mu\alpha} = -\frac{1}{2}
        \frac{F_{,\varphi}}{F}\bar{T}\bar{g}^{\alpha\mu}
        \bar{\nabla}_{\mu}\varphi\,.
  \label{eq:barT}
\end{equation}
From now on, we restrict our attention
to spherically symmetric, time independent
stellar models. More specifically, we employ
polar slicing and radial gauge in the Einstein frame,
so that the line element is of the form
\begin{equation}
  \mathrm{d}\bar{s}^2 = \bar{g}_{\mu\nu}\mathrm{d}x^{\mu}\mathrm{d}x^{\nu} =
        -F\,\alpha^2 \mathrm{d}t^2+F\,X^2 \mathrm{d}r^2 + r^2\mathrm{d}
        \Omega^2\,,
  \label{eq:lineelement}
\end{equation}
where $\alpha$ and $X$ as well as the scalar field $\varphi$
are functions of the radius $r$, and $\mathrm{d}\Omega^2$ denotes the standard
line element on the unit 2-sphere. It is furthermore common practice to
introduce the gravitational potential $\Phi$ and mass function $m$ according to
\begin{equation}
  F\,\alpha^2 = e^{2\Phi},~~~~~F\,X^2 = \left(1-\frac{2m}{r}\right)^{-1}\,.
  \label{eq_nu_m}
\end{equation}
In this work we explore the behaviour of NSs in equilibrium and model
their matter as a perfect fluid at zero temperature; the temperature
of NS interiors in equilibrium,
despite being of order $10^{6}\,\mathrm{K}$,
is well below the Fermi temperature
$\mathcal{O}(10^{11})\,\mathrm{K}$ of matter at nuclear densities.
The energy momentum tensor is then given in terms of the
baryon density $\rho(r)$, the specific
enthalphy $h(r)$ and the pressure $P(r)$ by
\begin{eqnarray}
  &&\bar{T}_{\alpha \beta}=\frac{1}{F} T_{\alpha \beta}
        = \frac{1}{F} \left( \rho h u_{\alpha} u_{\beta} + P g_{\alpha \beta}
        \right)\,,
        \nonumber \\
  && \text{with}~~~~~~~~
        u^{\alpha} = \left[ \alpha^{-1},~0,~0,~0 \right]\,,
  ~~~~~~~~h = 1+\epsilon + \frac{P}{\rho}\,,
 \label{eq:mom}
\end{eqnarray}
and where $\epsilon$ is the specific internal energy.
Inserting the Einstein frame metric (\ref{eq:lineelement})
and the energy momentum tensor (\ref{eq:mom}) into
the field equations (\ref{eq:barG})-(\ref{eq:barT}),
we obtain the set of differential equations
\begin{eqnarray}
  \partial_r \Phi &=& \frac{F X^2-1}{2r}+\frac{4\pi rP X^2}{F}
        +\frac{r X^2 \eta^2}{2}-Wr X^2 F\,,
        \label{eq:Phi}\\
  \partial_r X &=& \frac{4\pi r X^3}{F}(\rho h -P)+\frac{r X^3 \eta^2}{2}\,,
        -\frac{X^3 F}{2 r}+\frac{X}{2 r}
        -\frac{F_{,\varphi} X^2 \eta}{2 F}+X^3FWr\label{eq:X}\,,\\
  \partial_r P &=& -\rho hF X^2\left(\frac{m}{r^2}+4\pi r \frac{P}{F^2}
        +\frac{r}{2F}{\eta}^2-rW\right)+\rho h \frac{F_{,\varphi}}{2F}X
         \eta\,,
        \label{eq:P} \\
  \partial_r \varphi &=& X\eta \label{eq:varphi}\,,
        \\
  \partial_r \eta &=& -\frac{3 \eta}{2 r}
        -\frac{2 \pi X F_{,\varphi}}{F^2}(\rho h-4P)
        -\frac{X^2 \eta F}{2 r}-\frac{4 X^2 \eta \pi r P}{F}
        -\frac{X^2 \eta^3 r}{2}
        \nonumber\\
  && +\frac{X \eta^2 F_{,\varphi}}{2 F}+X^2\eta FWr+XF\partial_\varphi W\,.
        \label{eq:eta}\,
\end{eqnarray}
In order to close this set of equations, we need to relate the pressure
and internal energy to the baryon density. In this work, we use cold polytropic
EOSs with exponent $\Gamma$,
\begin{equation}
  P = K \rho^{\Gamma}\,,
  \label{eq:EOS}
\end{equation}
The internal energy is then determined by the
first law of thermodynamics $dE=\dbar Q-pdV$ for adiabatic processes
with $\dbar Q=0$. For a total baryon number $N$
and mass per baryon $m_b$, the specific internal
energy and baryon density are given by
$\epsilon = E/(Nm_b)$ and $\rho = m_b N/V$, respectively, and the first law
results in
\begin{equation}
  \epsilon = \frac{P}{(\Gamma-1)\rho}\,.
  \label{eq:epsilon}
\end{equation}

The set of equations (\ref{eq:Phi})-(\ref{eq:eta}) can then be solved
subject to the boundary conditions
\begin{eqnarray}
  \text{at}~~r=0:~~~~\,&&\eta = 0\,,~~~~~~~~~~~FX^2 = 1\,,
        \nonumber \\
  \text{at}~~r\rightarrow \infty:~~&&\Phi = 0\,,~~~~~~~~~~
        \rho = 0\,,~~~~~~~~~~\varphi = 0\,.
  \label{eq:BCs}
\end{eqnarray}
The computation of solutions to this problem is complicated by three issues,
which we list in increasing order of difficulty.
\begin{itemize}
\item [(i)] The boundary conditions are specified at different locations of the
domain, so that we have a {\em two-point-boundary-value problem}.
\item [(ii)] For realistic values of the polytropic exponent $\Gamma$,
the pressure will reach zero at a finite radius $R_S$; at this
point, we need to match to an exterior solution with vanishing
baryon density $\rho$.
\item [(iii)] The asymptotic behaviour of the
scalar field near infinity is determined by the scalar mass $\mu$ and is given by
\begin{equation}
  \lim_{r\rightarrow\infty} \varphi \sim A_1 \frac{e^{-(\mu/\hbar) r}}{r}+A_2
        \frac{e^{(\mu/\hbar) r}}{r}\,,
  \label{eq:scalarasymptotics}
\end{equation}
for constants $A_1$, $A_2$. We are only interested in bounded solutions
with $\varphi \propto e^{-(\mu/\hbar) r}/r$.
This exponential fall-off
is responsible for the suppressed scalar contribution in the
interaction of pulsar binaries in massive ST gravity and forms
the key motivation for our study. From a purely numerical
point of view, however, Eq.~(\ref{eq:scalarasymptotics})
creates a significant challenge. Numerical algorithms
will pick up all possible modes of a solution --
even if only through roundoff error.
\end{itemize}
We therefore seek an algorithm
that provides us with explicit control over the asymptotic behaviour
of our numerical solutions.
In the next section, we will discuss how
this can be achieved inside the more standard frameworks employed to
address items (i) and (ii) of our above list.

\section{Numerical framework}
\label{sec:numerics}
Numerical methods for solving two-point-boundary-value problems
are well developed and fall into two main classes, {\em shooting
algorithms} and {\em relaxation schemes} (including collocation methods) \cite{Press1992}
To the best of our knowledge, all literature on
static NS models in ST gravity has employed
shooting algorithms; see e.g.~\cite{Ramazanoglu:2016kul,Yazadjiev:2016pcb}.
This process integrates the differential equations from one end of the domain
by supplementing the known boundary conditions at this point with
appropriate trial values for the remaining variables. The resulting integration
will typically not match the boundary conditions at the other end of the
domain, but the degree of violation can be used, e.g.~through a
Newton-Raphson or a bisection method, to iteratively improve the
trial values until all boundary conditions are satisfied within
a user-specified threshold accuracy.

For the case of our system of differential equations
(\ref{eq:Phi})-(\ref{eq:eta}) with boundary conditions
(\ref{eq:BCs}), this would work as follows. We first note
that the function $\Phi$ appears only in Eq.~(\ref{eq:Phi})
and in the form of its spatial derivative. 
We can therefore set $\Phi(0)=0$ and add an arbitrary
constant to match its boundary condition
{\em after} having solved for all variables.
Bearing in mind this freedom, we start the integration
at the origin $r=0$ by selecting values for the central baryon density
$\rho(0)$, the central scalar field amplitude $\varphi(0)$ and the
metric function $\Phi(0)$, additionally to the known $\eta(0)=0$
and $X(0)=1/\sqrt{F(\varphi(0))}$. The integration will reach
zero pressure at a finite $r$ which represents the NS radius. Beyond
this radius, the integration continues setting $\rho=P=0$ in
Eqs.~(\ref{eq:Phi})-(\ref{eq:eta}). In principle this surmounts the
issue (ii) mentioned in the previous section. We note, though, that
$P=0$ is, in general, not realized on a grid point which adds
a small discontinuity to the solution; the data on the outermost
grid point inside the star and on the first point outside
the star do not satisfy Eqs.~(\ref{eq:Phi})-(\ref{eq:eta}).
This discontinuity is
typically not problematic, but we will see below how it can be
simply eliminated in a relaxation approach.
For a massless scalar field, it is even possible to analytically
match the spacetime to an exterior
vacuum\footnote{The term ``vacuum'' here refers to the baryonic matter;
the scalar field is nonzero exterior to the star.}
metric; cf.~Eqs.~(8), (9) in \cite{Damour:1993hw}.
Integration beyond the neutron star radius is not required and the trial value $\varphi(0)$ can be improved
in accordance with the selected shooting algorithm.

Such an analytic matching is not known, however, for
{\em massive} scalar fields. And now
a more problematic issue arises as the integration is continued
beyond the NS radius; no matter how accurate the central value
$\varphi(0)$ has been chosen, the numerical solution will
contain an exponentially growing contribution
from the asymptotic behaviour (\ref{eq:scalarasymptotics})
and eventually blow up exponentially. Worse, this blowup prevents us from
improving our trial value $\varphi(0)$ through measuring the departure
from the correct boundary condition at infinity; this departure is
infinite and, hence, useless for numerical purposes.
In shooting algorithms, this problem is circumvented by
imposing the outer boundary conditions at a finite radius rather
than infinity; cf.~Sec.~III A in \cite{Ramazanoglu:2016kul}.

While this method is still capable of generating accurate stellar models,
a scheme covering the complete exterior and imposing the boundary conditions
at infinity provides practical advantages besides the more rigorous
boundary treatment. By extending all the way to infinity,
our scheme can provide initial data for time evolutions on arbitrarily large
computational domains (including compactified evolution schemes that
incorporate spatial or null infinity) without resorting to adhoc
procedures to extend results beyond the inevitably finite
blow-up radius of shooting methods.
We will also obtain a vacuum exterior that is matched to the NS
interior on a grid point; in fact, the value of the NS
surface, rather than the central density,
will select the specific stellar model. Furthermore, the relaxation
scheme provides an exceptionally elegant and simple way to implement
the matching between the interior and exterior domain that we
expect to be applicable to a wider range of problems, including
extension to time evolutions.

For our method, we first introduce the NS radius $R_S$
as a free parameter.
On the domain $r\in[0,R_S]$, we use the differential equations
(\ref{eq:Phi})-(\ref{eq:eta}) with boundary conditions (\ref{eq:BCs})
for $\eta$ and $X$ at $r=0$; the condition $FX^2=1$ is formulated as
an equation involving the unknown $\varphi(0)$. In the exterior,
we introduce a compactified radial coordinate
\begin{equation}
  y = \frac{1}{r}\,,
\end{equation}
set $\rho=P=0$, and introduce rescaled variables
\begin{equation}
  \sigma = \varphi e^{(\mu/\hbar)r}\,,~~~~~~~~~~
  \kappa = -\eta e^{(\mu/\hbar)r}\,.
\end{equation}
By factoring the exponential dependence into our scalar field
variables, we ensure that regular solutions $\sigma$ and $\kappa$
asymptote towards a $\propto y$ dependence at $y=0$. We find this
step crucial in achieving convergence of our relaxation scheme which
struggles with the exponential fall-off of $\varphi$ and $\eta$
but copes smoothly with the benign linear behaviour of the
rescaled $\sigma$ and $\kappa$. We also notice a minor (but not crucial)
improvement in the speed of convergence when switching from $X$ to the mass
function $m$ of Eq.~(\ref{eq_nu_m}) and hence use the set of variables
$\Phi,~m,~\sigma,~\kappa$ in the exterior. The differential
equations in the exterior domain $y\in [0,y_{\rm S}]$,
$y_{\rm S}=1/R_{\rm S}$, thus become
\begin{eqnarray}
  \partial_y \Phi &=& -\frac{m}{1-2my}-\frac{1}{2\left(1-2my\right)
        y^3e^{2\mu/y}} \left(\frac{\kappa^2}{F}-\mu^2\sigma^2\right)\,,
  \label{eq:Phiy}
  \\
  \partial_y m &=& -\frac{1}{2y^4e^{2\mu/y}}\left(
        \frac{\kappa^2}{F}-\mu^2\sigma^2\right)\,,
  \label{eq:my}
  \\
  \partial_y \sigma &=& \frac{X\kappa-\mu\sigma}{y^2}\,,\\
  \partial_y \kappa &=& -\kappa\partial_y\Phi+\frac{FX\mu^2\sigma+\kappa
        \left(2y-\mu\right)}{y^2}+\frac{X\kappa^2}{2y^2}
        \frac{F_{,\varphi}}{F}\,,
  \label{eq:kappay}
\end{eqnarray}
and the matching conditions imposed at the surface of the NS are given by
\begin{eqnarray}
  \Phi(y_{\rm S}) &=&\Phi(R_{\rm S})\,,
  \nonumber \\
  m(y_{\rm S}) &=& \left[ \frac{r}{2}\left(1-\frac{1}{FX^2}\right)
        \right]_{r=R_{\rm S}}\,,
  \nonumber \\
  \sigma(y_{\rm S}) &=& \varphi(R_{\rm S}) e^{(\mu/\hbar) R_{\rm S}} \,,
  \nonumber \\
  \kappa(y_{\rm S}) &=& -\eta(R_{\rm S}) e^{(\mu/\hbar) R_{\rm S}}\,.
  \label{eq:matching}
\end{eqnarray}
We formally also use the trivial equation $\partial_y P=0$ in the exterior
which allows us to use a constant number of five variables over the entire
grid. This grid consists of $N$ grid points in the interior and $M$ points
in the exterior. We discretize the differential equations using
cell-centered second-order stencils which provides us with $5(N-1)$
algebraic equations in the interior and $5(M-1)$ equations in the exterior.
The boundary conditions provide two further equations at $r=0$ and
three further equations at $y=0$. The surface radius is represented
twice on our grid, the outermost point $r=R_{\rm S}$ of the interior
and the innermost point $y_{\rm S}$ of the exterior grid. The variables
used on these points are related by the matching conditions
(\ref{eq:matching}) as well as the trivial $P(R_{\rm S})=P(y_{\rm S})=0$.
In total, we thus have $5(N+M)$ non-linear algebraic equations for the
$5(N+M)$ unknown values of the variables on the grid points.
Given an initial guess, we can linearize the equations around this
trial solution which leads to a matrix equation with block-diagonal
structure that is readily inverted to improve the guess iteratively;
we use the algorithm of Ref.~\cite{Press1992} and typically obtain
convergence after about ten iterations. The initial guess is obtained
by integrating Eqs.~(\ref{eq:Phi})-(\ref{eq:eta}) up to $R_{\rm S}$,
fixing $\sigma$ and $\kappa$ as linear functions $\propto y$
in the exterior and integrating Eqs.~(\ref{eq:Phiy}), (\ref{eq:my})
with these specified scalar sources. Note that in this calculation
we set $\rho=P=0$ in the exterior irrespective of whether or not they
have reached zero at $R_{\rm S}$; the discontinuity that may result
at the matching point is removed in the ensuing relaxation process.

Even for modest resolutions such as $N=M=401$, this approach provides
an accuracy of $O(10^{-4})$.
All models discussed in the remainder of this
paper have been computed with this code.


\section{Results}
\label{results}
%
%
\subsection{Overall phenomenology}
We start this section by defining the terminology
and diagnostic quantities as well as
providing a qualitative review of the
different branches of static NS models encountered in massive ST gravity.
We then explore in the following subsections in more detail the
impact of the ST parameters $\alpha_0$, $\beta_0$ and $\mu$ on the
structure of these branches.

In the following, we will use the term
``family'' for the set of all NS models obtained for fixed
EOS and ST parameters $\alpha_0$, $\beta_0$ and $\mu$.
We will use the term ``branch'' to denote a subset of solutions of a
family that share some specific property, for example strong
scalarization. A family thus consists of one or more branches.
In some cases, we find a branch to have the shape of
a closed loop disconnected from other branches, and we also
refer to such a branch as a ``loop''.
For reference, we note that the scalar mass
$\mu$ introduces a Compton wavelength and characteristic frequency
given by
\begin{equation}
  \lambda_C = 1.24\times 10^6\,\mathrm{km}~\left(
  \frac{\mu}{10^{-15}\,\mathrm{eV}}\right)^{-1}\,,~~~~~~~~~~
  f_* = \frac{\omega_*}{2\pi} = 24.2\,\mathrm{Hz}~\left(
  \frac{\mu}{10^{-15}\,\mathrm{eV}}\right)^{-1}\,.
\end{equation}

Unless stated otherwise, our numerical NS models in massive
ST theory are computed with the
polytropic EOS labelled ``EOS1'' in Ref.~\cite{Novak:1997hw}. Translated
into our notation,
we therefore compute the pressure and specific internal energy
from the baryon density $\rho$ through
Eqs.~(\ref{eq:EOS}) and (\ref{eq:epsilon}) with
\begin{equation}
  K
  =1.543~\frac{\mathrm{cm}^{3\Gamma-1}}{\mathrm{g}^{\Gamma-1}\mathrm{s}}
  ,~~~~~~~~~~ \Gamma=2.34\,.
  \label{eq:EOS1}
\end{equation}
The families of solutions thus obtained are conveniently represented in
a mass-radius diagram. For this purpose, we define the total
baryon mass $M_b$ as the volume integral of the baryon number density
multiplied by the mass per baryon $m_b$. Translated into our
baryon density $\rho = m_b n_b$, the expression becomes
\begin{equation}
  M_b=m_b\int d^3 x\sqrt{-g}n_bu^t=4\pi\int_{0}^{R_S}dr
        \left(r^2\frac{\rho}{F^{3/2}\sqrt{1-2m/r}}\right)\,.
\end{equation}
The motivation for using the baryon mass, rather than the gravitational
mass of Eq.~(\ref{eq_nu_m}), arises from the conservation of the baryon
number; if we consider the possibility that a NS might migrate
dynamically from one branch to another, we expect it to do so
at constant $M_b$, whereas the binding energy and, hence, gravitational
mass will, in general, change.

\begin{figure}[t]
  \centering
  \includegraphics[width=0.48\textwidth]{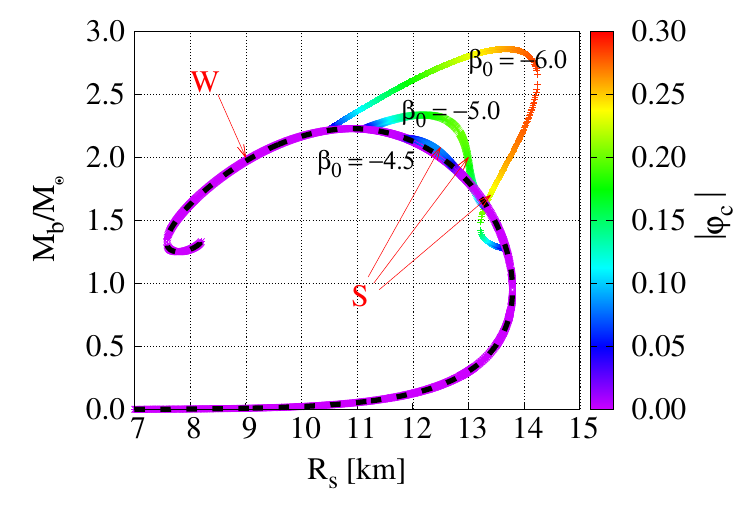}
  \includegraphics[width=0.48\textwidth]{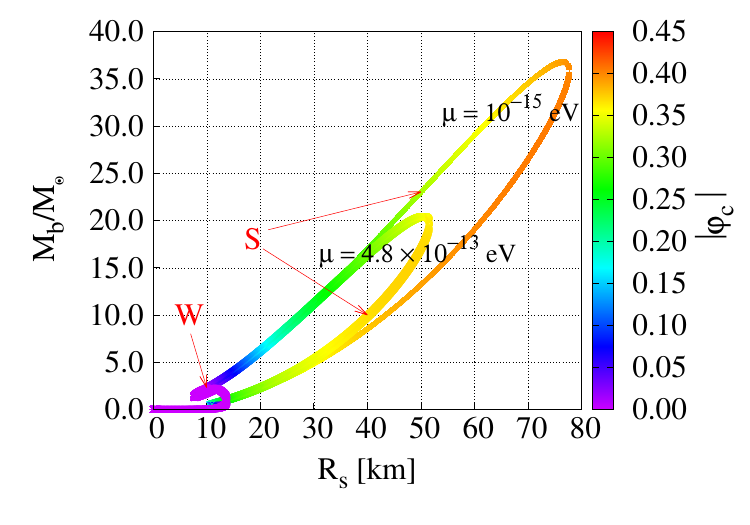}
  \caption[$M_b-R$ plots for varying $\beta_0$ and $\mu$.]
          {Branches of NS models are shown in the form of
           baryon-mass vs.~radius ($M_b-R_S$) diagrams.
           {\em Left:} For fixed values $\alpha_0=10^{-4}$
           and $\mu=4.8\times 10^{-13}\,\mathrm{eV}$,
           we plot the strongly scalarized branches obtained
           for selected values of $\beta_0$.
           For reference, the dashed black curve displays the
           solutions obtained in GR with
           $\alpha_0=\beta_0=0$.
           {\em Right:} Here we fix $\alpha_0=10^{-4}$
           and a more extreme value of $\beta_0=-15$ and vary the scalar mass $\mu$; larger deviations from the GR structure are clearly visible in this case.
           In both panels the color scale measures the central value of
           $|\varphi|$ and the ``S'' and ``W'' label the strongly and weakly scalarized branches described in the text.
          }
  \label{fig:various}
\end{figure}

As in massless ST theories, all NS solutions can be classified as
either {\em weakly scalarized} with scalar field profiles reaching
a magnitude $\varphi\sim O(\alpha_0)$ or {\em strongly scalarized}
solutions where the scalar field reaches values
$\varphi\sim O(1)$ \cite{Damour:1993hw}.
In this work, we call these branches W (for weak)
and S (for strong scalarization); see e.~g.~Fig.~\ref{fig:various}.
The distinction between the two regimes naturally blurs for large values
$\alpha_0 = \mathcal{O}(1)$; in this work we consider only $\alpha_0 \ll 1$
and thus retain a clear division between weakly and strongly scalarized
NSs.

As is well known, the NS solutions in GR\footnote{We produce GR solutions
with our code by simply setting $\alpha_0=\beta_0=0$.} form a
one-parameter family whose members can be characterized by its
central density $\rho_c$ \cite{Shapiro1983}. For the EOS (\ref{eq:EOS1}),
this leads to the black branch in the left panel\footnote{The smooth curves in all our mass-radius plots are in fact made
up of a large number of crosses which in most cases are not individually
visible. We opt {\it not} to connect these with lines to avoid spurious
cross-branch connections. In consequence, some curves appear to have
breaks when the gradient becomes nearly vertical; these breaks are not
physical.} of Fig.~\ref{fig:various}.
In ST gravity, the introduction of a scalar field leads to additional
branches of NS solutions with a non-vanishing scalar-field profile
$\varphi(r)$. The shape of the extra branches depends on the
values of the parameters $\alpha_0$, $\beta_0$ and $\mu$.
In agreement with the literature \cite{Novak:1998rk},
we observe strongly scalarized
solutions if $\beta_0 \lesssim -4.35$.
These new branches are displayed in the left panel of
Fig.~\ref{fig:various}
in terms of a color code that
denotes the central scalar field amplitude. In contrast, the
weakly scalarized solutions we obtain for $\beta_0 \gtrsim -4.35$
have macroscopic properties that barely differ from those of the
GR solutions and their branch would be indistinguishable from the GR
family in the figure. For these cases, we have set
$\alpha_0=10^{-4}$ and $\mu = 4.8\times 10^{-13}\,\mathrm{eV}$. We now
discuss in more detail how the solutions and their branches vary when
the parameter values are changed.

\subsection{Dependence on \texorpdfstring{$\mu$}{mu}}
\label{sec:mu}
The variation of scalarized NS branches in massive ST theory has
already been studied in Ref.~\cite{Ramazanoglu:2016kul} who
generally observe that an increase in the scalar mass $\mu$
results in a weakening of the scalarization. By computing
a sequence of models with equal gravitational or
``Arnowitt-Deser-Misner'' (ADM) \cite{Arnowitt:1962hi} mass,
they observe a monotonic decrease in the scalarization as
the scalar mass is increased. Around $10^{-12}\,\mathrm{eV}$,
their scalar profile drops to negligible levels
when $\beta_0=-4.5$; cf.~their Fig.~2.
Our results exhibit a similar drop in scalarization. We
illustrate this general behaviour
by comparing the cases $\mu=10^{-15}\,\mathrm{eV}$
and $\mu=4.8\times 10^{-13}\,\mathrm{eV}$ in the right
panel of Fig.~\ref{fig:various}; the color code of the branches
represents the central scalar field amplitude and displays lower
values for the larger $\mu$.

We furthermore notice that the strongly scalarized NS branches
reach out to smaller baryon mass and radii as we increase
the scalar mass $\mu$. As we shall discuss in
more detail in Sec.~\ref{sec:stability}, the stable NS model
for a given baryon mass is that with the largest radius.
For fixed $M_b$, a larger scalar mass $\mu$ 
results in smaller and more compact stable NS models.

\subsection{Dependence on \texorpdfstring{$\alpha_0$}{alpha0}}
\label{sec:alpha0}
\begin{figure}[t]
  \centering
  \includegraphics[width=0.48\textwidth]{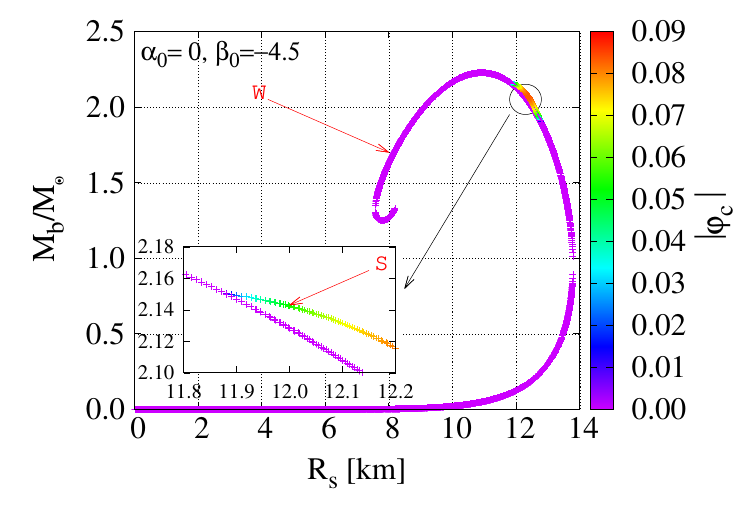}
  \centering
  \includegraphics[width=0.48\textwidth]{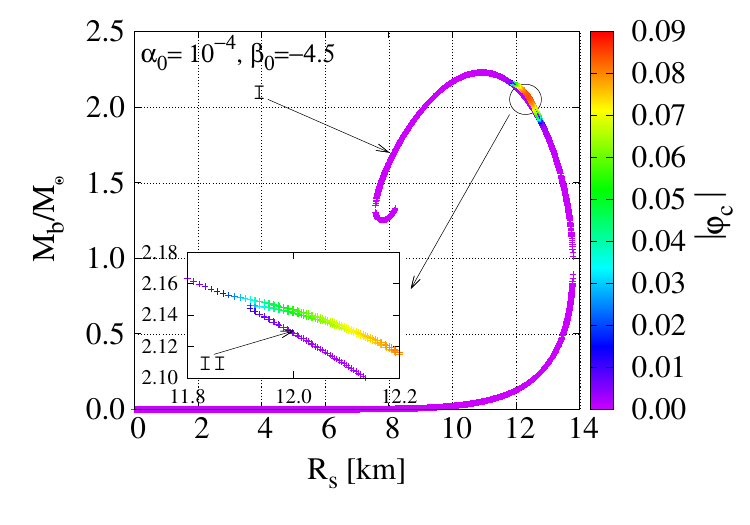}
  \centering
  \includegraphics[width=0.48\textwidth]{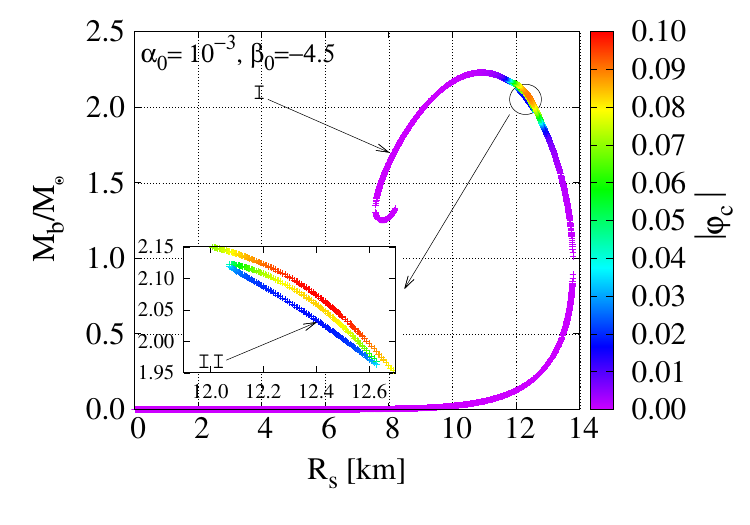}
  \caption[$M_b-R$ plots for varying ${\alpha}_0$ and $\beta_0=-4.5$.]
          {$M_b-R_S$ diagrams are shown for
          $\mu=4.8\times 10^{-13}\,\mathrm{eV}$ and $\beta_0=-4.5$,
          as well as $\alpha_0=0$ (\textbf{top left}),
          $\alpha_0=10^{-4}$ (\textbf{top right}) and $\alpha_0=10^{-3}$
          (\textbf{bottom panel}).
          The color scale measures the central value of $|\varphi|$.
          Whereas the S and W branches connect at two points when
          $\alpha_0=0$, the S branch splits in two for nonzero $\alpha_0$
          with each part connecting to GR-like models in such a way
          that we obtain a ``loop'' of models separate from the
          main branch of solutions. We refer to the main branch as
          branch I and to the loop as branch $II$.
    }
  \label{split_starting}
\end{figure}

\begin{figure}[t!]
  \centering
  \includegraphics[width=0.48\textwidth]{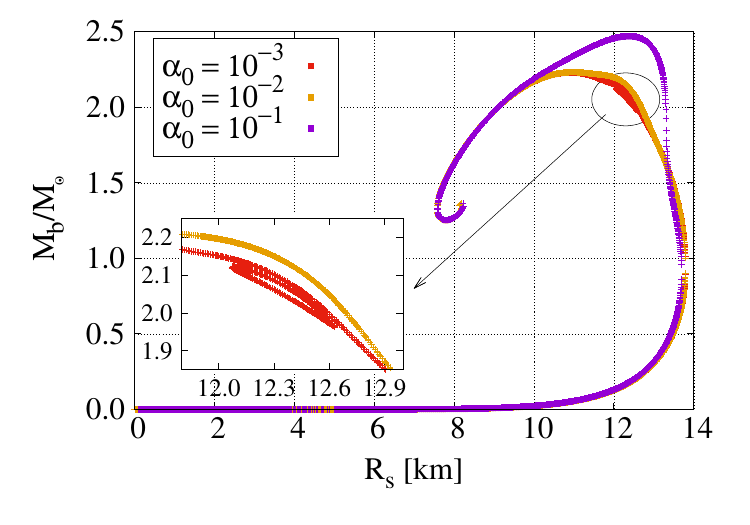}
  \centering
  \includegraphics[width=0.48\textwidth]{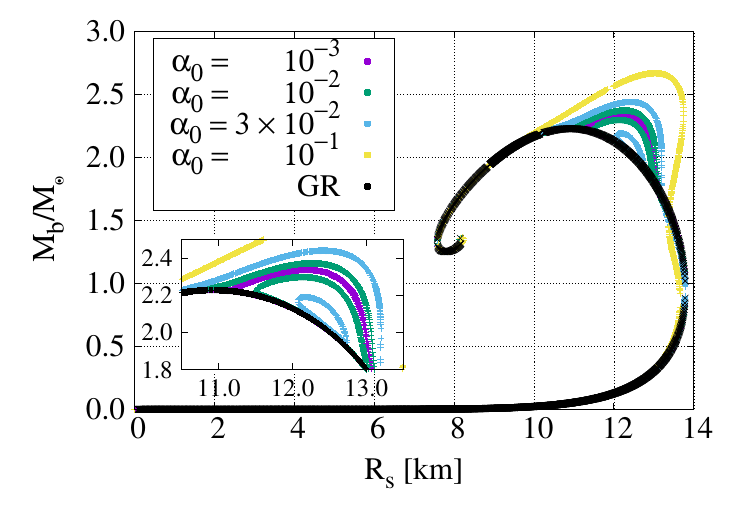}
  \caption[$M_b-R$ plots for varying $\alpha_0$.]
          {$M_b-R_S$ diagrams are shown for a scalar mass $\mu=4.8\times 10^{-13}$
           with $\beta_0=-4.5$ (left) and $-5$ (right).
           As we increase $\alpha_0$, the loop of branch $II$ solutions
           shrinks in size and eventually disappears. For reference,
           we include the GR branch in the right panel.
          }
  \label{closed_contour}

\bigskip \bigskip

  \centering
  \includegraphics[width=.49\linewidth]{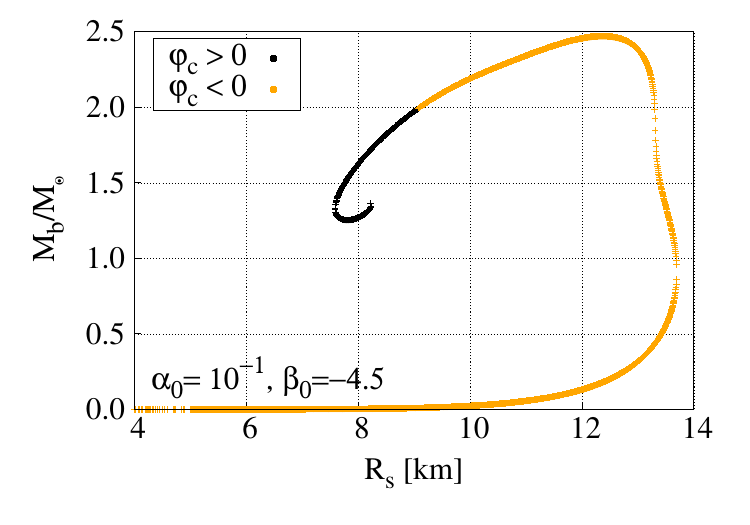}
  \includegraphics[width=.49\linewidth]{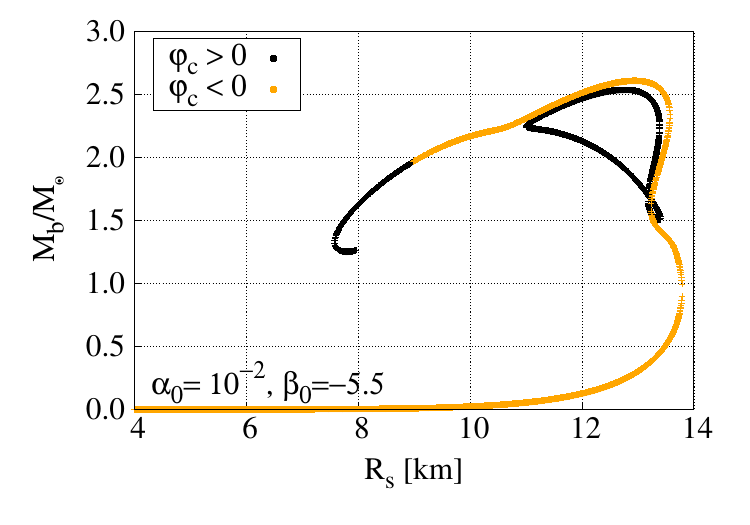}
  \\
  \includegraphics[width=.49\linewidth]{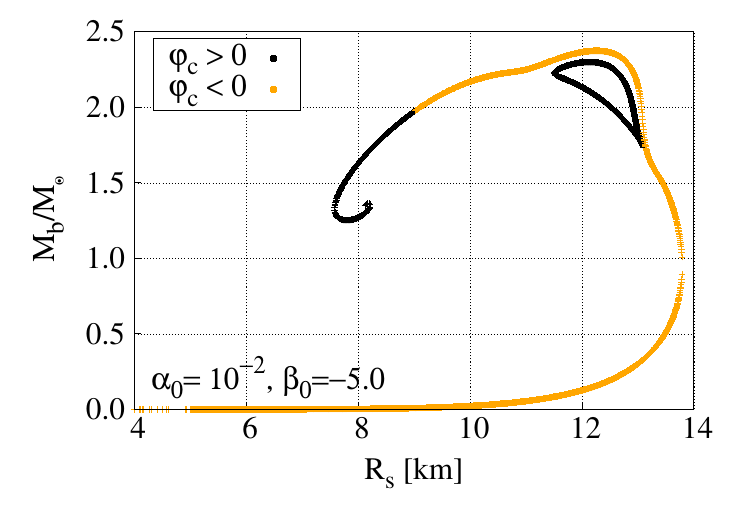}
  \includegraphics[width=.49\linewidth]{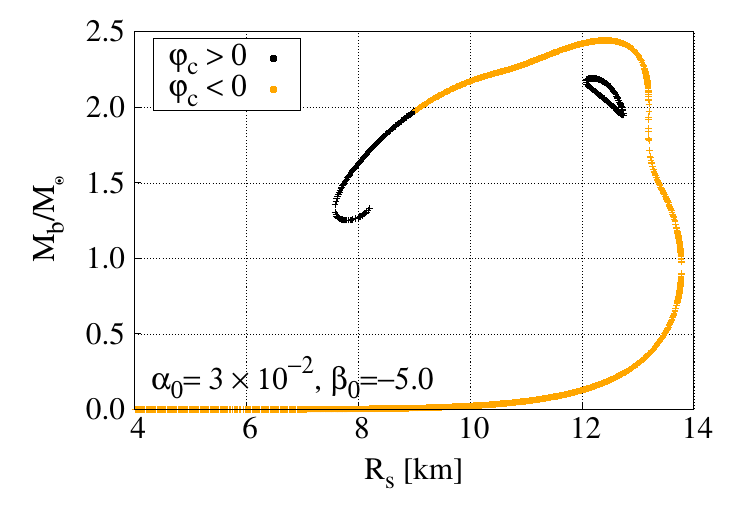}
  \caption[Sign of central scalar field of NSs for several ST parameters.]
       {Distribution of scalarized NS models based on the sign
        of $\varphi_c$ in the $M_b-R_S$ plane for
        $\mu=4.8\times 10^{-13}\,\mathrm{eV}$ and, from top-left
        to bottom-right, $(\alpha_0,\beta_0)=
        (10^{-1},-4.5)$,  $(10^{-2},-5.5)$,~~~
         $(10^{-2},-5)$, $(3\times 10^{-2},-5)$.
        The orange points represent models with $\varphi_c<0$ whereas the
        black ones have $\varphi_c>0$.
        The type $II$ models on the loop differ in the sign of $\varphi_c$
        from the nearby main branch $I$ models. Furthermore,
        we always observe a sign
        flip at the high-density end of branch $I$ (around
        $R_S\approx 8\,\mathrm{km}$ in the figure) but these NS models
        are unstable; cf.~Sec.~\ref{sec:stability}.
        }
        \label{positive_negative_1}
\end{figure}

When $\alpha_0=0$, our system of equations (\ref{eq:Phi})-(\ref{eq:eta})
is invariant under the transformation\footnote{Recall that
$\varphi \to -\varphi$ implies $\eta\to -\eta$ and that
$F_{,\varphi}$ and $V_{,\varphi}$ are linear in $\varphi$ when
$\alpha_0=0$.}
$\varphi \to - \varphi$. In this case, each strongly scalarized model
consists of two solutions that only differ by a minus sign in the
scalar-field profile. Additionally to this degenerate scalarized branch,
there exists a branch with zero scalar field, i.e.~the set of models
we also obtain in GR.

A nonzero $\alpha_0$ breaks the degeneracy of the strongly scalarized
branch which now splits into two branches with unequal
macroscopic properties and whose scalar field magnitudes
differ at a level $\mathcal{O}(\alpha_0)$. This split is illustrated
in Fig.~\ref{split_starting} where we consider NS models in the
$M_b-R_S$ plane for $\mu = 4.8\times 10^{-13}\,\mathrm{eV}$,
$\beta_0=-4.5$ and different value of $\alpha_0$.
For $\alpha_0=0$, we see that the scalarized branch directly connects
to the GR branch. For $\alpha_0\ne 0$, we obtain a weakly scalarized branch W
with $\varphi = \mathcal{O}(\alpha_0)$ in place of the GR models.
In terms of their mass $M_b$ and radius $R_S$, however, these models
are barely distinguishable from their GR counterparts, and we refer to
them as ``GR like'' models. The strongly
scalarized branch S, on the other hand, splits into two, each of them
connecting to separate parts of the GR-like branch in such a way
that we obtain one loop of models that is separated from the
single, large branch; cf.~the insets in the top right and bottom
panels of Fig.~\ref{split_starting}. We find the gap between the
loop and the main branch to be proportional to $\alpha_0$ and
independent of $\beta_0$.
For small but nonzero $\alpha_0$, we thus find two sets of solutions:
A branch $I$ that approximately follows the $M_b-R_S$ curve of GR
for small and for large central density $\rho_c$, and a separate
branch $II$ on a closed loop located in the region where
strong scalarization occurs. The central baryon density strictly increases
as we move along branch $I$, starting at $(M_b,R_S)=(0,0)$.
We note that branches $I$ and $II$ both contain weakly as
well as strongly scalarized models. Instead, their distinction arises from
the separation of the loop from the main branch of models.

As we increase $\alpha_0$, however, the loop of branch $II$ solutions
shrinks and eventually disappears, leaving branch $I$ as the only
class of solutions. This remaining branch approximately overlaps
with the GR family in the very high and low $\rho_c$ regime but
shows a strong bulge of strongly scalarized solutions when
$\rho_c$ has values comparable to nuclear density.
In agreement with the literature, we observe that these
scalarized models can reach significantly larger masses and radii than
their GR counterparts.
We illustrate these features in Fig.~\ref{closed_contour}
for $\beta_0=-4.5$ and $-5$, but note that this behaviour occurs
universally in all cases we have studied.

We conclude the discussion of the $\alpha_0$ dependence with a subtle
observation we make throughout our computations: For all NS models of
branch $II$ the central scalar field value has the same sign;
$\varphi_c>0$ for our convention. For the vast majority (though not all)
branch $I$ solutions, $\varphi_c$ has the opposite sign; $\varphi_c<0$ in
our case. We display this observation graphically in
Fig.~\ref{positive_negative_1}
for several combinations of $\beta_0$ and $\alpha_0$.
We note, however, that branch $I$ always contains a swap in
$\mathrm{sign}(\varphi_c)$ at very large central baryon density:
we always observe $\varphi_c>0$ as $\rho_c\rightarrow\infty$.

\begin{figure}[t]
  \centering
 \includegraphics[width=0.60\linewidth]{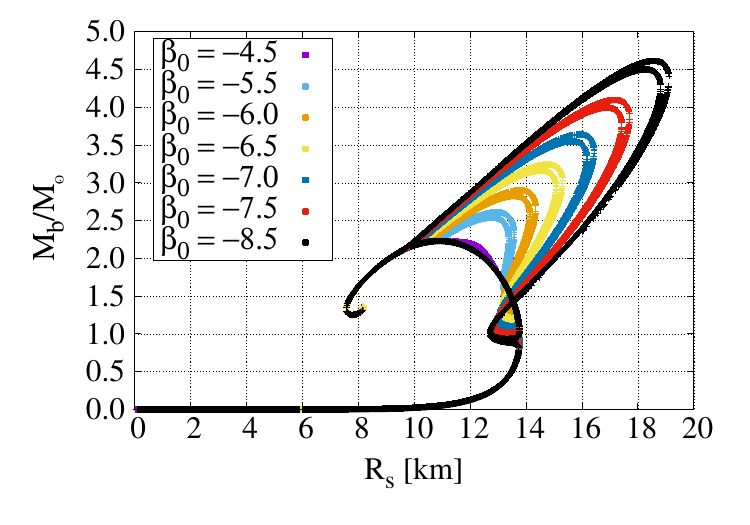}
 \caption[$M_b-R$ plots when varying $\beta_0$ in the presence of a
          large $\alpha_0$ parameter.]
         {$M_b-R_S$ diagrams are shown for several values of $\beta_0$ in the
          regime of spontaneous scalarization $\beta_0<-4.35$.
          The other scalar field parameters are
          $\mu=-4.8\times 10^{-13}$ eV, $\alpha_0=10^{-2}$.
          For increasingly negative values of $\beta_0$, the
          S branch extends to larger values of the NS radius and
          baryon mass.
          }
  \label{various_alpha_0_01}
\end{figure}

\subsection{Dependence on \texorpdfstring{$\beta_0$}{beta0}}
\label{sec:beta_dependence}
\begin{figure}[t]
  \centering
  \includegraphics[width=0.48\textwidth]{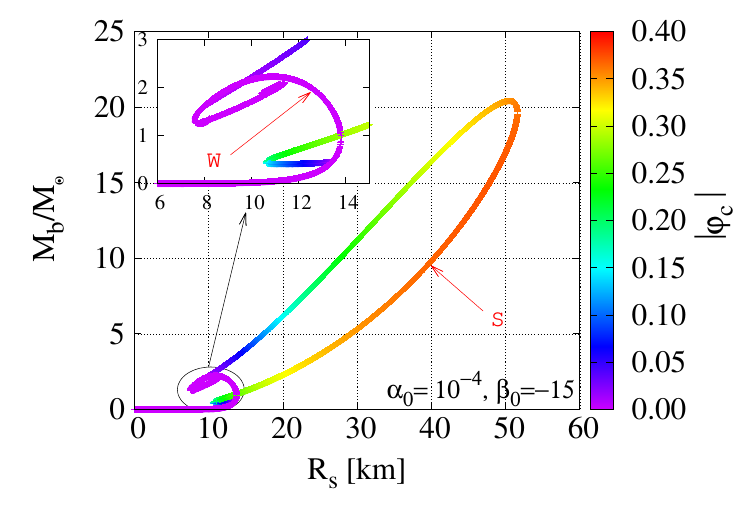}
  \centering
  \includegraphics[width=0.48\textwidth]{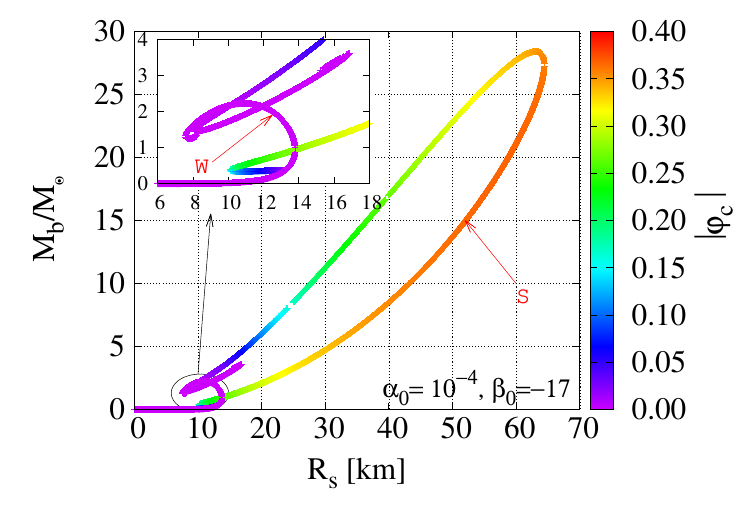}
  \centering
  \includegraphics[width=0.48\textwidth]{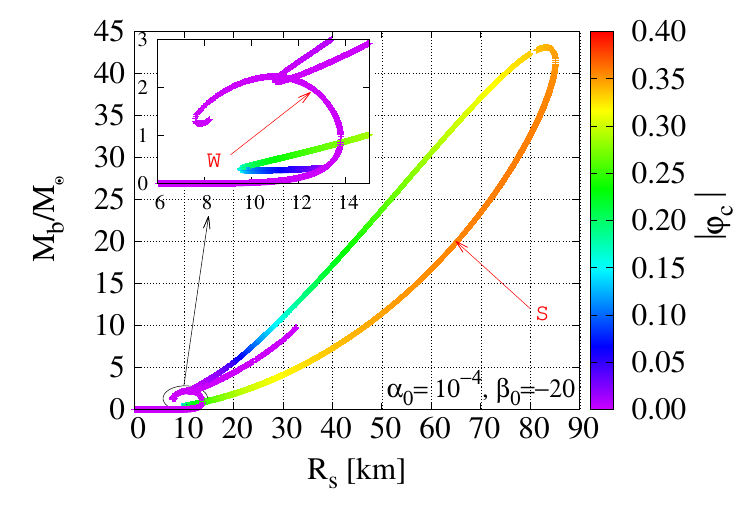}
  \centering
  \includegraphics[width=0.48\textwidth]{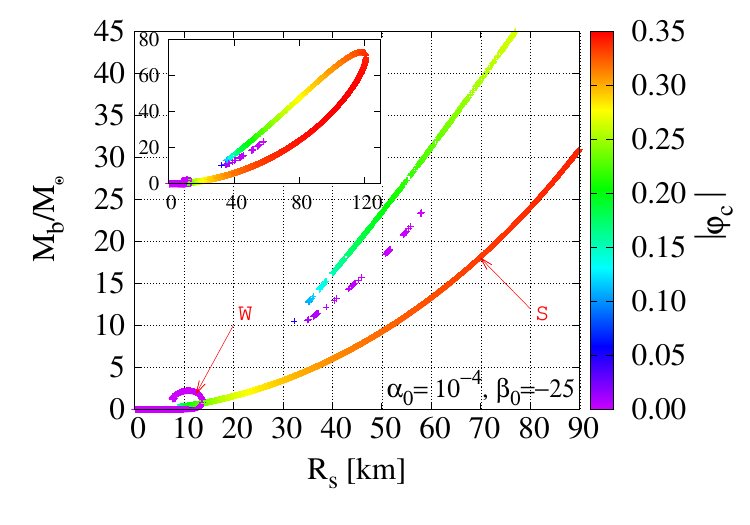}
  \caption[$M_b-R$ plots for very negative ${\beta}_0$.]
        {$M_b-R_S$ diagrams are shown for
         $\mu=4.8\times 10^{-13}\,\mathrm{eV}$ and $\alpha_0=10^{-4}$,
         as well as $\beta_0=-15$ (\textbf{top left}),
         $\beta_0=-17$ (\textbf{top right}),
         $\beta_0=-20$ (\textbf{bottom left})
         and $\beta_0=-25$ (\textbf{bottom right panel}).
         The color scale measures the central value of $|\varphi|$.
         This sequence of plots (from top left to bottom right) shows the upper end of the S branch disconnecting and separating from the W branch as $\beta_0$ becomes more negative.
        }
  \label{plot_within_plot}
\end{figure}

Spontaneous scalarization is a non-linear phenomenon and driven by the
quadratic coupling parameter when $\beta_0 \lesssim -4.35$.
It has already been remarked in \cite{Ramazanoglu:2016kul}, that
this threshold value for strong scalarization is barely affected
by the introduction of a non-zero scalar mass. Our results confirm
this observation as is illustrated by the W and S branches of solutions
shown in the left panel of Fig.~\ref{fig:various} for $\beta_0=-4.5$,
$-5$ and $-6$ with fixed $\alpha_0=10^{-4}$ and $\mu=4.8\times 10^{-13}\,\mathrm{eV}$.
Note that the transition from weak to strong scalarization along a
sequence of NS models, while strictly speaking continuous, is
sufficiently abrupt to allow for a clear distinction between models belonging to branch W or S.

The three branches displayed in the left panel of
Fig.~\ref{fig:various} for $\beta_0=-4.5$, $-5$ and $-6$ also
demonstrate the increasing deviation in terms of mass and radius
of branch S models from their GR-like counterparts. Increasingly negative
values of $\beta_0$ allow for larger maximum mass and radius;
cf.~also Sec.~IV A in \cite{Ramazanoglu:2016kul}. This rather strong
effect may provide opportunities for constraining $\beta_0$
through mass and radius measurement of NSs, although a
reevaluation of the measurements in the framework of ST gravity
(rather than assuming GR) will be required for this purpose.
In Fig.~\ref{fig:various}, the strongly scalarized branch S models
appear as an arc splitting off from the GR-like branch W. We always find
branch S to have this qualitative shape and the size of the
arc grows monotonically as $\beta_0$ takes on increasingly negative
values. This is illustrated in Fig.~\ref{various_alpha_0_01},
which displays branches W and S for several values of
$\beta_0 \le -4.5$. This figure also demonstrates that branch S
has a shape resembling an inverted 'S'; it initially splits off
from branch W towards {\em smaller} radii (around
$R_S\approx 13.5\,\mathrm{km}$ and $M_b \approx 1\,M_{\odot}$ in
the figure) before turning around and crossing branch W towards
larger $R_S$. Note also that for each choice of $\beta_0$,
branch S consists of two nearby but distinct curves; this splitting
results from the relatively large value $\alpha_0=10^{-2}$ as we have
already seen in the last section; cf.~the bottom right panel in
Fig.~\ref{positive_negative_1}.

For highly negative $\beta_0$, we obtain NS models with yet larger
radius and baryon mass as illustrated in Fig.~\ref{plot_within_plot},
where we plot branches W and S for $\beta_0=-15$,
$-17$, $-20$, $-25$
and fixed
$\alpha=10^{-4}$, $\mu=4.8\times 10^{-13}\,\mathrm{eV}$.
In this figure, we also notice a new effect:
the upper (in the sense of larger
$\rho_c$) end of branch S exhibits a more complex
structure. Instead of connecting to branch W, branch S appears to
remain separate and curl around; cf.~the insets for
$\beta_0=-15$ and $\beta_0=-17$. This behaviour becomes
clearer for yet more negative $\beta_0$: between
$\beta_0=-20$ and $\beta_0=-25$, the intersection of
branch S with branch W is lost and instead, branch S
forms its own tail of NS models with very small
central values of the scalarfield $|\varphi_c|$;
note the magenta color of this end of branch S. Contrary to
what one might guess, the NS models on this tail of branch S
are still {\em strongly} scalarized; the profile
$\varphi(r)$ merely reaches its maximum away from the center
$r=0$.

\begin{figure}[t!]
  \includegraphics[width=0.49\textwidth]{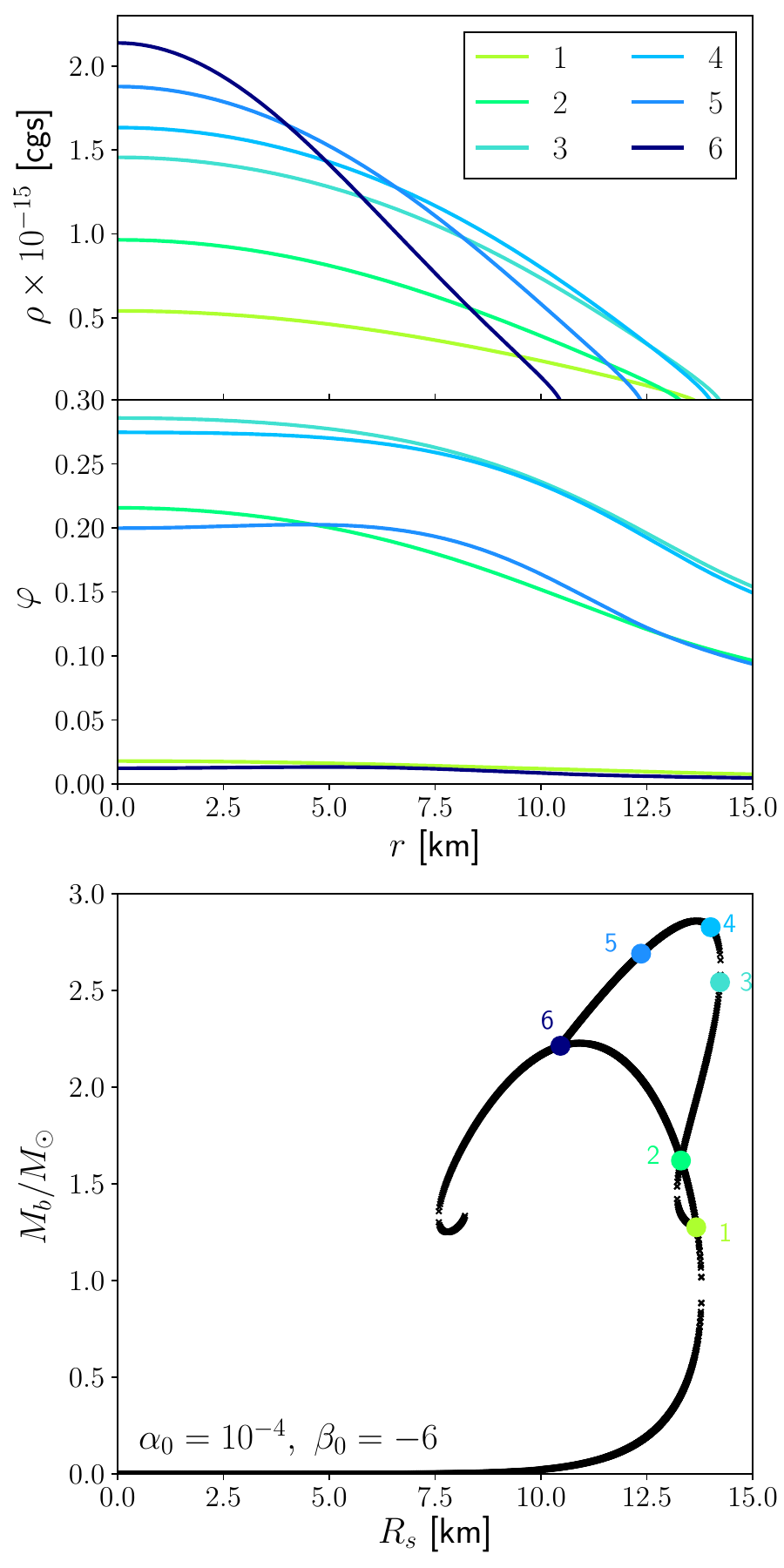}
  \includegraphics[width=0.49\textwidth]{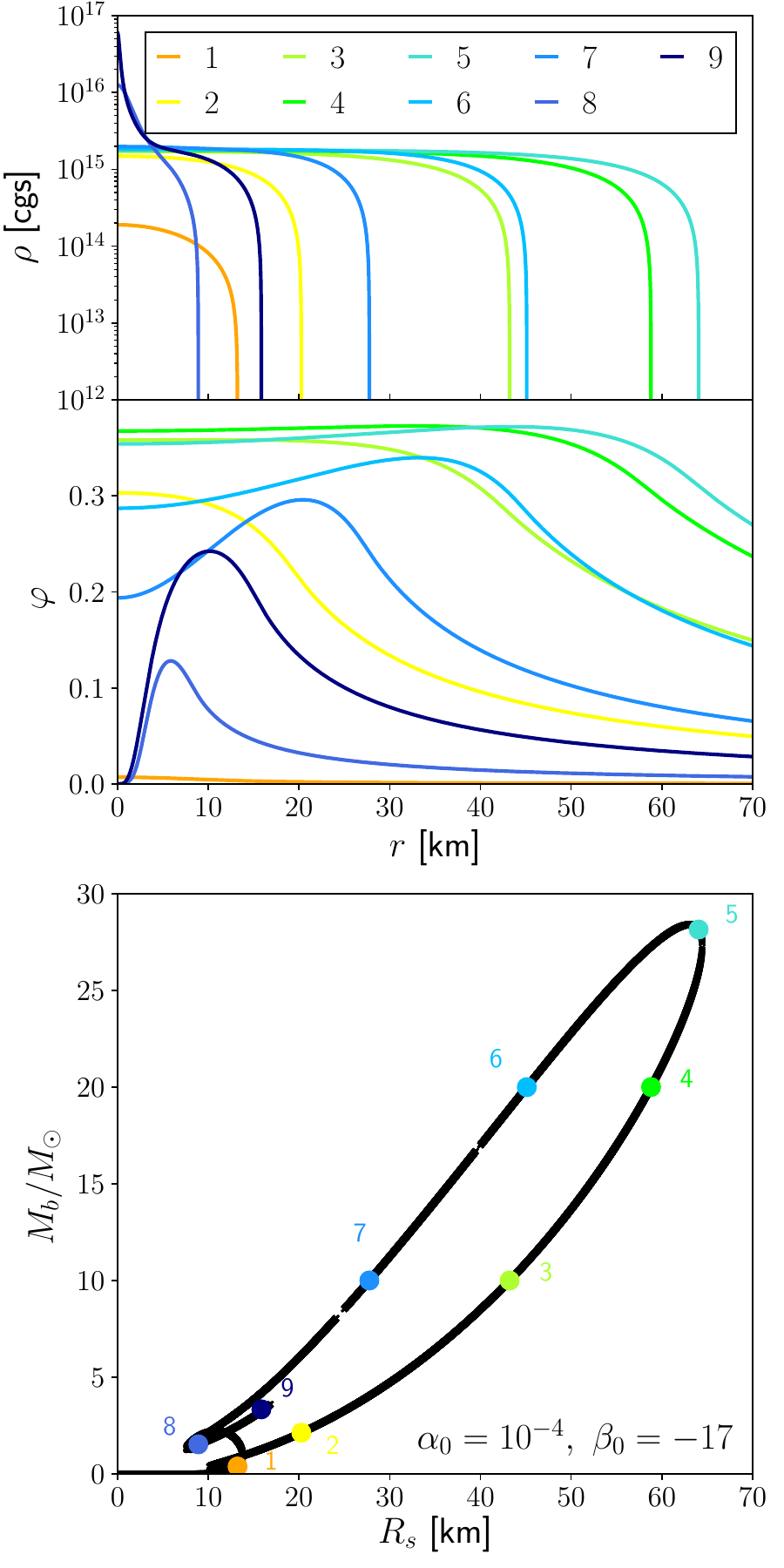}
  \caption[Baryon density and scalar field profiles for several
           strongly scalarized solutions.]
       {The branches of NS models are shown in the $M_b-R_S$
        plane in the bottom panels
        for $\mu=4.8\times 10^{-13}\,\mathrm{eV}$,
        $\alpha_0=10^{-4}$ as well as
        $\beta_0=-6$ (left) and $\beta_0=-17$ (right).
        Several NS models are marked along the branches as
        colored circles. The top panels show
        the radial profiles of the baryon density $\rho(r)$
        and the scalar field $\varphi(r)$ for these
        NSs using their respective color.
        The density profile always reaches a maximum at the origin; however, the scalar field profile in some cases reaches a peak at a non-zero radius.
        }
  \label{fig:frames}
\end{figure}

We explore this behaviour in more detail by comparing
in Fig.~\ref{fig:frames} the
sequence of models obtained for $\beta_0=-6$ with that for
$\beta_0=-17$, keeping $\alpha_0=10^{-4}$ and
$\mu=4.8\times10^{-13}\,\mathrm{eV}$ fixed. The bottom
panels show the respective families in the $M_b-R_S$ diagram
analogous to Fig.~\ref{plot_within_plot}. Along the
branches, we have marked several NSs by colored circles,
and for these models we plot the baryon-density and scalar-field
profiles $\rho(r)$, $\varphi(r)$ in the upper
panels\footnote{We have selected here exclusively models with
$\varphi_c>0$. The small $\alpha_0=10^{-4}$ leads to such a
small splitting that the corresponding figure using the
models with $\varphi_c$ would be indistinguishable besides
the sign reversal in $\varphi(r)$.}.
For $\beta_0=-6$, we observe a simple pattern: At the lower
branch point, we obtain a weakly
scalarized model (``1'') with comparatively small central
baryon density. As we continue along branch S, the scalar
field increases in strength, reaching a maximum at
maximal radius (model ``3''). Beyond that point, the
central baryon density $\rho_c$ keeps increasing, but the
scalar profile weakens. Eventually (model ``6''),
a weakly scalarized but highly compact NS marks the smooth
reconnection with branch W; note that this weakly scalarized
model is located on the unstable branch of the GR-like models.

The analogous results for $\beta_0=-17$ in
the right column of Fig.~\ref{fig:frames} display
a qualitatively similar behaviour near the lower branch point
(model ``1''); the central baryon-density and scalar-field
values increase as we move along branch S (model ``2'').
Eventually, however, the scalar field profile changes its
qualitative behaviour and peaks away from $r=0$ while
the central value $\varphi_c$ decreases. In consequence,
the upper tail of branch S now consists of models
with $\varphi_c \approx 0$ but strong scalarization
at $r>0$ and does not directly connect to branch W;
compare model ``8'' for $\beta_0=-17$ with model ``6'' for $\beta_0=-6$. As branch S curls around,
the scalar profile strengthens once again and we encounter
models with a steep density cusp (model ``9'').
We cannot rigorously rule out that after further curling around, branch
S might eventually connect with branch W, but our numerical results
do not show any signs of this happening.
We finally note the remarkable structure of these upper tail
stars: A highly compact star of baryonic matter is surrounded
by a shell of scalar-field (i.e.~bosonic) matter. This structure
is reminiscent of the atom like shape noticed for
stars in massless ST gravity in \cite{Brito:2015yfh} and
scalarized black holes in modified gravity~
\cite{Baumann:2019eav}.

\subsection{Stability of models}
\label{sec:stability}
\begin{figure}[t]
  \centering
  \includegraphics[width=.49\linewidth]{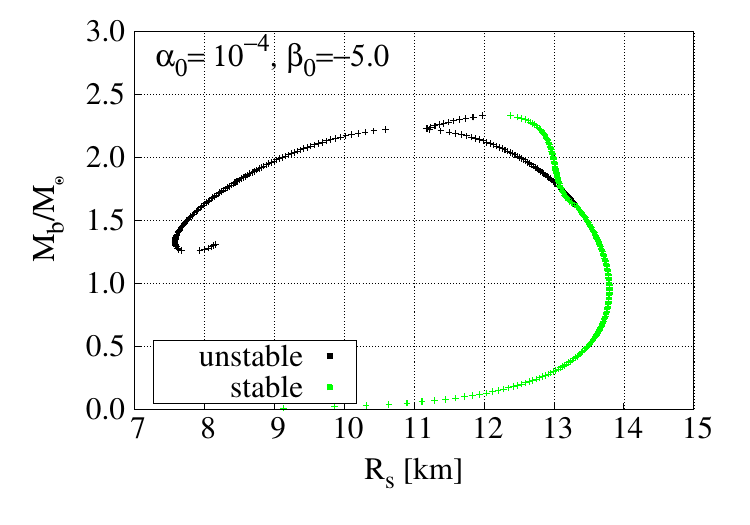}
  \includegraphics[width=.49\linewidth]{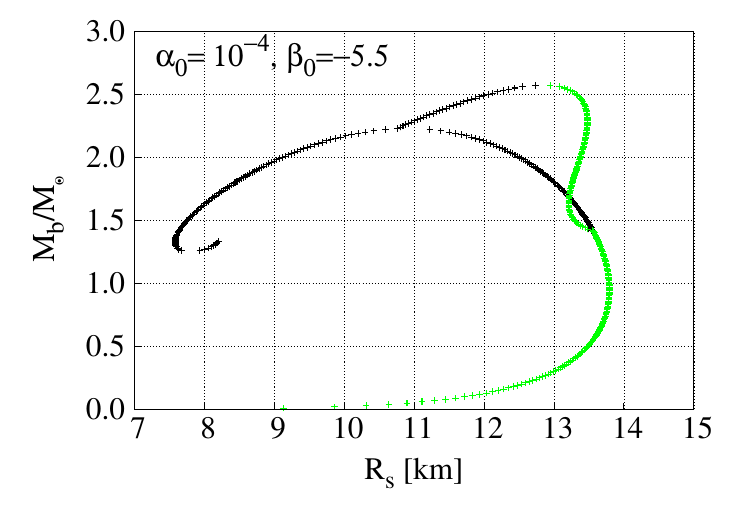}
  \includegraphics[width=.49\linewidth]{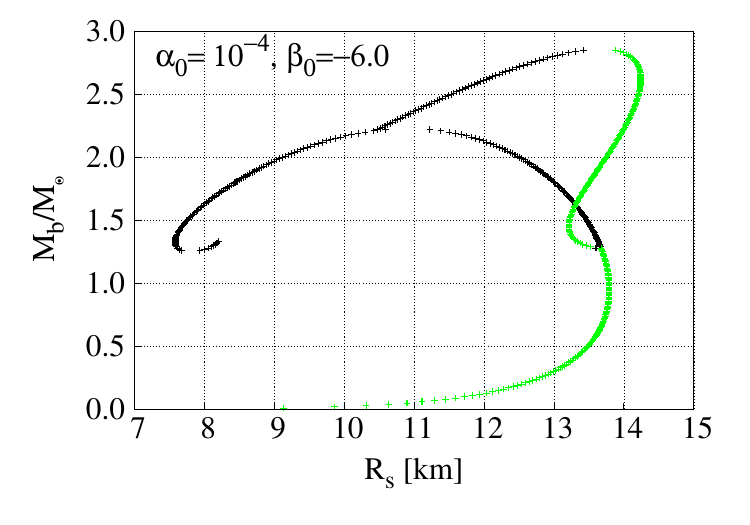}
  \includegraphics[width=.49\linewidth]{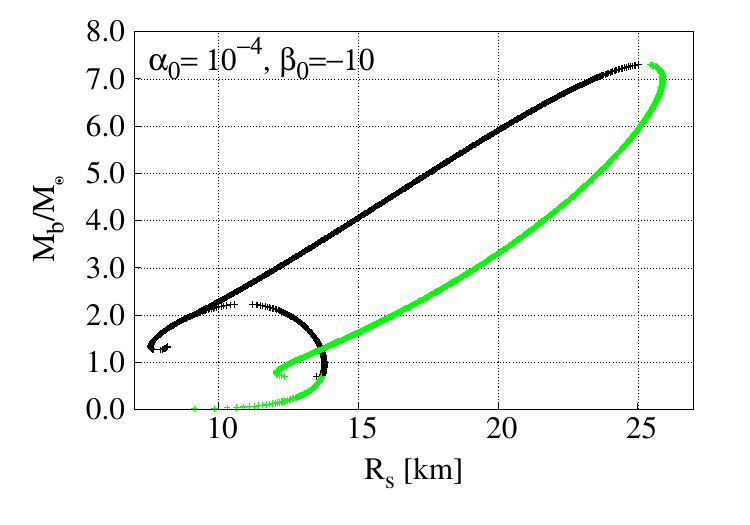}
  \caption[Stability of scalarized NSs with very negative $\beta_0$ values.]
        {Plots showing the distribution of stable (green) and unstable (black) 
         NS configurations in the $M_b-R_S$ plane.
         When two solutions with the same baryon mass $M_b$ exist, the one with the lower ADM mass is energetically favored.
         The scalar parameters are $\mu=4.8\times 10^{-13}$ eV,
         $\alpha_0=-10^{-4}$ and, from top left to bottom right,
         $\beta_0=-5$, $-5.5$, $-6$ and $-10$.
         }
  \label{stable models}
\end{figure}
\begin{figure}[t]
  \centering
  \includegraphics[width=.49\linewidth]{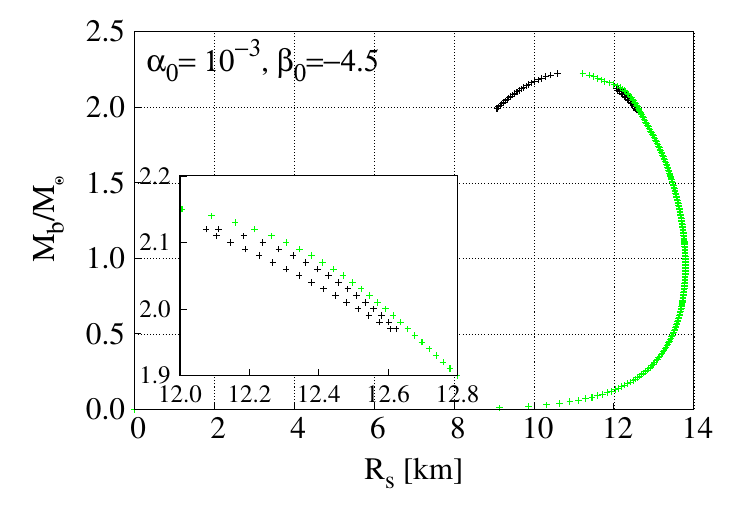}
  \includegraphics[width=.49\linewidth]{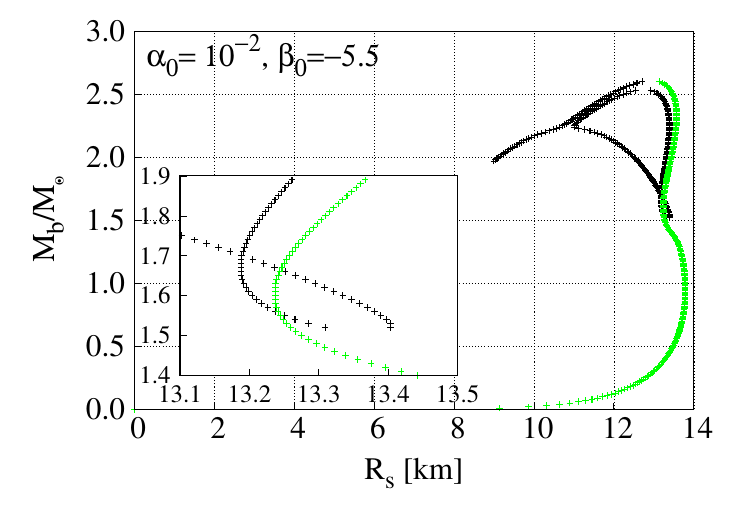}
  \includegraphics[width=.49\linewidth]{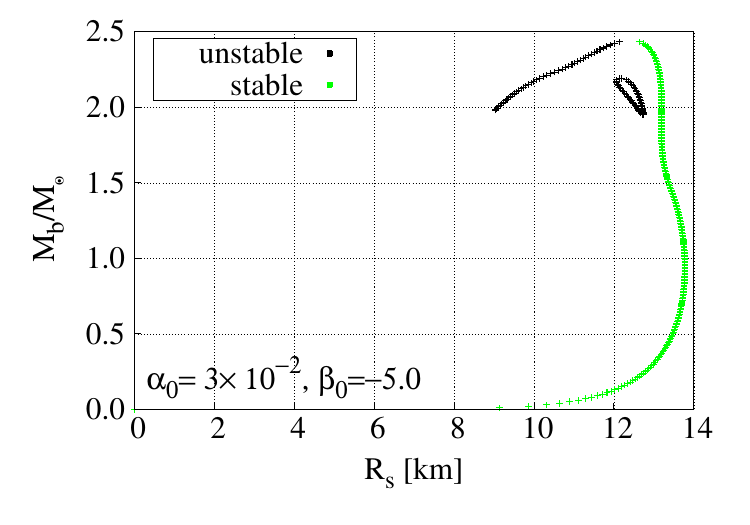}
  \caption[Stability of scalarized NSs with very large $\alpha_0$ values.]
        {Same as Fig.~\ref{stable models} using scalar mass
         $\mu=4.8\times 10^{-13}$ eV, and coupling parameters
         from top-left to bottom, $(\alpha_0,\beta_0)=(10^{-3},-4.5)$,
         $(10^{-2},-5.5)$, $(3\times10^{-2},-5)$.
         }
  \label{stable models_split}
\end{figure}
In the previous sections, we have seen many cases where for
fixed ST parameters $\alpha_0$, $\beta_0$, $\mu$ several
equilibrium NS models with equal baryon mass exist; see
e.g.~$M_b = 2\,M_{\odot}$ in the left panel of
Fig.~\ref{fig:various}. We can analyze the stability of these
models by comparing their binding energy. The model with the lowest
ADM mass, i.e.~strongest binding energy, represents the stable configuration
and other models with equal baryon mass are expected to migrate
to this configuration under perturbations.
We note, however, that the physical relevance of this
instability depends on the instability timescale as compared
to other dynamical timescales under consideration.

Using this method, we classify in Fig.~\ref{stable models} stable and
unstable NS models for several values of $\beta_0$ at fixed
$\alpha=10^{-4}$ and $\mu=4.8\times 10^{-13}\,\mathrm{eV}$.
The results confirm the theoretical prediction that the
weakly scalarized branch W becomes unstable when strongly
scalarized counterparts with equal baryon mass exist
\cite{Novak:1998rk}. Note also that the scalarized
branch S exhibits a stability structure analogous to the
well known GR case: The maximum mass model separates
stable from unstable stars and the stable models are those
with larger radius.

In Fig.~\ref{stable models_split}, we analyze how the stable
and unstable models spread among our ``loop'' branches
$I$ and $II$ of Sec.~\ref{sec:alpha0} for different values
of $\alpha_0$. The stable NSs are the strongly scalarized
models with the largest radius, whereas the NSs on branch
$II$ (i.e.~on the closed loop) are always unstable.
As a general pattern in all our computations, we find the
models with the strongest central scalar field value $|\varphi_c|$
to be the stable configurations. For our convention, these
always turn out to be models with $\varphi_c<0$;
for example compare Fig.~\ref{stable models_split}
with Fig.~\ref{positive_negative_1}.

\section{Conclusions}
\label{sec:conclusions}
In this study, we have numerically computed solutions of spherically
symmetric NSs in massive ST theory using a numerical scheme that
enables us to eliminate the exponentially growing modes from the
scalar field. For this purpose, we split the domain into the NS
interior and the exterior from the stellar surface to infinity
and discretize the resulting equations with a second-order
relaxation scheme. This method enables us to compute NS spacetimes extending
all the way to infinity where we can prescribe the boundary conditions
in simple Dirichlet form. This formalism also provides a trivially
simple implementation of the matching conditions without the need
to perform interpolation.

We have used the resulting code to compute solutions of static,
spherically symmetric NSs in massive ST theory and explored in
detail the structure of the resulting branches of solutions in
the (baryon) mass-radius plane for combinations of the
linear and quadratic coupling parameters $\alpha_0$, $\beta_0$ of
the ST theory and the scalar mass $\mu$. We summarize the main findings
of our analysis as follows.
\begin{itemize}
  \item In agreement with previous literature studies of NS equilibrium
        models in massive and massless ST gravity, we find
        larger values of $\alpha_0$ and $\beta_0$ to result
        in larger deviations from the NS solutions in GR,
        whereas larger values of the scalar mass tend to reduce
        these deviations; cf.~Figs.~\ref{fig:various}
        and \ref{closed_contour}.
  \item For $\alpha_0=0$, the NS models of GR are also solutions
        of the field equations of massive ST gravity. For
        $\beta\lesssim -4.35$, we find, additionally
        to the GR branch, the spontaneously scalarized
        class of NS solutions that Damour \& Esposito-Far{\`e}se
        discovered in their original exploration of massless
        ST theory \cite{Damour:1993hw} and that were
        also identified in massive ST theory in
        \cite{Ramazanoglu:2016kul}. These solutions are invariant
        under the scalar field transformation $\varphi \rightarrow
        -\varphi$.
  \item A non-zero $\alpha_0$ breaks this degeneracy and results in a dissection
        of the branches around the branch points; instead of the two
        connected branches of scalarized and non-scalarized solutions
        for $\alpha_0=0$, we now find a main branch $I$ and a smaller
        loop of branch $II$ solutions; cf.~Fig.~\ref{split_starting}.
        The solutions on branches $I$ and $II$ are characterized by different
        signs of the central scalar-field value $\varphi_c$;
        cf.~Fig.~\ref{positive_negative_1}.
  \item For sufficiently negative $\beta_0$, roughly
        $\beta_0 \lesssim -15$, we observe a qualitative change
        in the strongly scalarized branch S of solutions. Instead
        of smoothly approaching the weakly scalarized branch W
        as happens for milder $\beta_0$,
        its upper (in the sense of increasing central baryon density) tail
        now either crosses or completely detaches from the
        W branch.
  \item For highly negative values of $\beta_0$, we
        furthermore encounter a new type of strongly scalarized
        solutions at this upper end of the S branch: the maximum of the scalar
        field is located away from the stellar center;
        cf.~Figs.~\ref{plot_within_plot}, \ref{fig:frames}.
        In its most extreme form, these solutions
        are composed of
        highly compact NS models surrounded by a scalar
        shell; see e.g.~\cite{Brito:2015yfh,Baumann:2019eav}
        for similar ``gravitational atom'' like configurations
        in other theories of gravity.
  \item Whenever multiple NS models with equal baryon mass exist,
        we find the scalarized model to be the stable configurations
        in the sense of minimal binding energy. Typically,
        though not always, this is the model with the largest
        radius; cf.~Figs.~\ref{stable models},
        \ref{stable models_split}. We also observe that the
        stable configurations
        agree in the
        sign of the central scalar field value, $\varphi_c<0$
        in our convention.
\end{itemize}
The behavior with respect to the scalar parameters seems to be
universal as we have encountered the same $M_b-R_S$ profile deviations
with respect to GR for all other equations of state that we have
studied.
We have explored in a similar manner, though less exhaustively, the cold hybrid EOS1, EOS3 and EOSa \cite{Rosca-Mead:2020ehn}, APR4 \cite{Akmal:1998cf}, 2H and HB \cite{Read:2009yp}
and observe qualitatively similar behaviour.

\vspace{6pt}




\funding{This research was funded by
the European Union’s H2020 ERC Consolidator
Grant “Matter and strong-field gravity: new frontiers in Einstein’s
theory” Grant No. MaGRaTh–646597 and the STFC Consolidator Grant
No. ST/P000673/1,
the GWverse COST Action Grant No. CA16104,
``Black holes, gravitational waves and fundamental physics''.
D.G. is supported by Leverhulme Trust Grant No. RPG-2019-350,
M.A. is supported by the Kavli Foundation,
Computational work was performed on the SDSC Comet and TACC Stampede2
clusters through NSF-XSEDE Grant No. PHY-090003; the Cambridge CSD3
system through STFC capital Grants No. ST/P002307/1 and No.
ST/R002452/1, and STFC operations Grant No. ST/R00689X/1; the
University of Birmingham BlueBEAR cluster; the Athena cluster at
HPC Midlands+ funded by EPSRC Grant No.~EP/P020232/1; and the
Maryland Advanced Research Computing Center (MARCC).
}

\acknowledgments{We thank Hector Okada da Silva, Fethi Ramazano\u{g}lu, and Christian Ott for discussions.}




\appendixtitles{no} 
\appendix



\reftitle{References}


\bibliography{regen}

\end{document}